\documentclass[pdflatex,sn-nature]{sn-jnl}% Single-column writing version
\usepackage{graphicx}%
\usepackage{multirow}%
\usepackage{amsmath,amssymb,amsfonts}%
\usepackage{amsthm}%
\usepackage{mathrsfs}%
\usepackage[title]{appendix}%
\usepackage{xcolor}%
\usepackage{textcomp}%
\usepackage{manyfoot}%
\usepackage{booktabs}%
\usepackage{tabularx}%
\usepackage{algorithm}%
\usepackage{algorithmicx}%
\usepackage{algpseudocode}%
\usepackage{listings}%
\usepackage{needspace}%
\usepackage{placeins}%
\usepackage{caption}%

\newcommand{\suppnote}[2]{%
  \par\vspace{0.8em}%
  \phantomsection%
  \label{suppnote:#1}%
  \noindent\textbf{Supplementary Note #1: #2}\par\vspace{0.5em}%
}

\makeatletter
\def\ps@silayout{%
  \let\@mkboth\@gobbletwo
  \let\@oddhead\@empty
  \let\@evenhead\@empty
  \def\@oddfoot{\hfil\thepage\hfil}%
  \let\@evenfoot\@oddfoot}
\makeatother

\let\SNsection\section
\renewcommand{\section}{\Needspace{12\baselineskip}\SNsection}
\let\SNsubsection\subsection
\renewcommand{\subsection}{\Needspace{9\baselineskip}\SNsubsection}
\let\SNabstractfont\abstractfont
\renewcommand{\abstractfont}{\SNabstractfont\unboldmath}
\makeatletter
\def\au@and{\global\advance\punctcount by -1\relax\ifnum\punctcount=1\ \&\else\unskip, \fi}
\makeatother
\theoremstyle{thmstyleone}%
\theoremstyle{thmstyletwo}%

\theoremstyle{thmstylethree}%

\begin{document}
% Replace the journal-template text block with the SI page geometry.
\newgeometry{top=2.2cm,bottom=2.2cm,left=2.4cm,right=2.4cm,
  bindingoffset=0mm,headheight=5.5pt,headsep=5.6mm,footskip=10mm}

\title[Conjugate phase-noise cancellation enables submicrometre dual-comb ranging with free-running megahertz-linewidth lasers]{Conjugate phase-noise cancellation enables submicrometre dual-comb ranging with free-running megahertz-linewidth lasers}

\author[1]{\fnm{Yue} \sur{You}}
\equalcont{These authors contributed equally to this work.}

\author[1]{\fnm{Zaifan} \sur{Wu}}
\equalcont{These authors contributed equally to this work.}

\author[1]{\fnm{Yi} \sur{Zou}}

\author[1]{\fnm{Xiang} \sur{Cai}}

\author[3]{\fnm{Yixiao} \sur{Zhu}}

\author[1]{\fnm{Chenbo} \sur{Zhang}}

\author[2]{\fnm{Dan} \sur{Lu}}

\author[5,6]{\fnm{Siming} \sur{Chen}}

\author[7]{\fnm{Shujie} \sur{Pan}}

\author[4]{\fnm{Xian} \sur{Zhou}}

\author[1]{\fnm{Xiaopeng} \sur{Xie}}

\author*[1,8]{\fnm{Fan} \sur{Zhang}}\email{fzhang@pku.edu.cn}

\affil[1]{\orgdiv{State Key Laboratory of Photonics and Communications, School of Electronics}, \orgname{Peking University}, \orgaddress{\city{Beijing}, \postcode{100871}, \country{China}}}

\affil[2]{\orgdiv{State Key Laboratory of Optoelectronic Materials and Devices}, \orgname{Institute of Semiconductors, Chinese Academy of Sciences}, \orgaddress{\city{Beijing}, \postcode{100083}, \country{China}}}

\affil[3]{\orgdiv{State Key Laboratory of Photonics and Communications, Department of Electronic Engineering}, \orgname{Shanghai Jiao Tong University}, \orgaddress{\city{Shanghai}, \postcode{200240}, \country{China}}}

\affil[4]{\orgname{University of Science and Technology Beijing}, \orgaddress{\city{Beijing}, \postcode{100083}, \country{China}}}

\affil[5]{\orgdiv{Laboratory of Solid State Optoelectronics Information Technology}, \orgname{Institute of Semiconductors, Chinese Academy of Sciences}, \orgaddress{\city{Beijing}, \postcode{100083}, \country{China}}}

\affil[6]{\orgdiv{College of Materials Science and Opto-Electronic Technology}, \orgname{University of Chinese Academy of Sciences}, \orgaddress{\city{Beijing}, \postcode{101804}, \country{China}}}

\affil[7]{\orgname{HS Photonics Co., Ltd.}, \orgaddress{\street{Xiangjiang Science \& Technology Innovation Base}, \city{Changsha}, \state{Hunan}, \postcode{413000}, \country{China}}}

\affil[8]{\orgname{Peng Cheng Laboratory}, \orgaddress{\city{Shenzhen}, \postcode{518055}, \country{China}}}

\abstract{Frequency-domain dual-comb ranging combines rapid acquisition with interferometric sensitivity, but high-performance implementations often rely on mutually coherent or actively stabilised comb sources. Free-running sources can reduce this hardware burden, but their phase noise and drift of the optical frequency offset can blur radio-frequency (RF) comb teeth and weaken probe--reference phase correlation. Previous phase-slope implementations have therefore relied on sufficiently resolved RF teeth, within-coherence-length probe--reference paths or explicit digital tracking of these fluctuations. Here we demonstrate a low-cost frequency-domain dual-comb ranging architecture that combines independent free-running distributed-feedback (DFB) lasers with conjugate phase-noise cancellation (CPNC). By forming a self-conjugate signal before the phases of individual RF-comb teeth are extracted, CPNC cancels the common laser phase factor and drifting optical-frequency-offset term while retaining the distance-dependent phase slope. Using electro-optic combs seeded by DFB lasers with linewidths of $12~\mathrm{MHz}$ and $9~\mathrm{MHz}$, we achieve an Allan deviation of $219~\mathrm{nm}$ at $246~\mu\mathrm{s}$ and reduce the single-frame distance standard deviation from $558~\mu\mathrm{m}$ to $9.31~\mu\mathrm{m}$ with CPNC. Across the tested megahertz-linewidth configurations, CPNC delivered minimum Allan deviations below $250~\mathrm{nm}$. These results show that CPNC enables submicrometre ranging in a low-cost frequency-domain dual-comb architecture with reduced source-stabilisation and phase-management complexity.}


\maketitle
\pagestyle{silayout}
\thispagestyle{silayout}

\clearpage
\section{Introduction}\label{sec1}

Precision dimensional metrology is increasingly important in semiconductor manufacturing and industrial inspection, where small errors must be detected without compromising throughput. Continued device scaling has tightened requirements for critical-dimension and feature-shape metrology \cite{orji2018metrology,diebold2018perspective}, while advanced device stacks impose increasingly stringent overlay-control requirements \cite{kim2009device,zhang2023xray}. Heterogeneous integration further increases structural complexity \cite{mahajan2026heterogeneous}. Related industrial applications require non-contact dimensional, surface and displacement measurements \cite{catalucci2022optical,berkovic2012optical}. These settings call for ranging systems that combine high precision, rapid acquisition and robust operation.

Laser-ranging methods encode distance through different observables, each with characteristic limitations \cite{amann2001laser}. Time-of-flight precision is constrained by detector and electronic timing resolution \cite{blais2004review,hallman2014detection,koerner2021models}; frequency-modulated continuous-wave ranging requires a broad, linear frequency sweep \cite{riemensberger2020massively,lukashchuk2024photonic}; and single-frequency interferometry is phase ambiguous. Optical frequency combs instead provide multiple mutually referenced frequencies, enabling distance to be encoded in the spectral phase \cite{minoshima2000high,vandenberg2015mode}.

Dual-comb ranging (DCR) uses two frequency combs with slightly different repetition rates to map optical timing and phase information into the radio-frequency (RF) domain \cite{coddington2016dual}. Through asynchronous optical sampling or multi-heterodyne detection, this Vernier mapping supports rapid acquisition within accessible electronic bandwidth while retaining interferometric phase sensitivity. Depending on the retrieval scheme, DCR can combine absolute ranging and high precision \cite{martin2022performance,zhao2018absolute}, while specialised schemes provide ultrarapid acquisition with an extended non-ambiguity range \cite{li2022ultrarapid}. These capabilities have been demonstrated using electro-optic combs \cite{weimann2018fast}, quantum-dash mode-locked laser diodes \cite{trocha2022ultrafast} and soliton microcombs \cite{trocha2018ultrafast,wang2020long,wang2025nanometric}. Dispersive-Fourier-transform retrieval provides another rapid fibre-comb implementation \cite{chang2024dispersive}. Collectively, reported systems span nanometre-to-micrometre precision, ultrafast acquisition, long-distance operation and chip-scale integration.

These performance gains, however, depend on preserving a usable phase relationship among the detected RF-comb teeth. For frequency-domain DCR, this requirement is particularly stringent because distance is inferred from the phase slope across individually resolved RF-comb teeth. Mutual coherence has been established using modulator-derived combs \cite{weimann2018fast} and counter-propagating solitons \cite{wang2025nanometric}. Repetition-rate or frequency locking provides active control \cite{yang2023micrometer,yan2025rapid}, while self-injection locking has enabled compact low-noise dual microcombs \cite{qin2025compact}. These strategies are highly effective but impose stringent coherence requirements on specialised comb generation and stabilisation. In microcomb implementations, this can involve high-Q resonator fabrication, controlled soliton operation, pump management and locking electronics. In addition to coherence management, acousto-optic frequency shifts have been used to position RF beat notes \cite{zhao2018absolute,weimann2018fast}, whereas serrodyne shifting provides an integration-oriented alternative \cite{guo2025serrodyne}. These source-generation and control requirements increase implementation complexity and restrict the use of low-cost free-running sources in frequency-domain DCR.

Free-running dual-comb operation has been explored in ranging and hyperspectral LiDAR \cite{yang2025free,camenzind2025broadband}, while software self-correction can reduce reliance on stabilisation hardware \cite{hebert2017self,yang2025free}. Time-domain envelope methods support absolute ranging but depend on temporal localisation and can be affected by pulse-shape distortion, detector noise and intensity fluctuations \cite{wang2022intensity,yang2025intensity}. Other approaches explicitly estimate the optical frequency offset, repetition-rate fluctuations or residual phase fluctuations \cite{hebert2017self,yang2025free}. Frequency-domain DCR has also been demonstrated with independent free-running quantum-dash mode-locked laser diodes of approximately $15~\mathrm{MHz}$ linewidth \cite{trocha2022ultrafast}. There, an approximately $497~\mathrm{MHz}$ RF-comb spacing preserved tooth resolvability, while a probe--reference path imbalance below the coherence length maintained phase correlation. Such a broad RF span increases detector and digitiser bandwidth demands, and the comb spacing of some source platforms cannot be freely enlarged. Existing approaches therefore place phase robustness in coherent-source hardware, restrictive spectral and path conditions, or explicit phase tracking. These constraints motivate a low-cost architecture that preserves phase robustness with free-running megahertz-linewidth sources.

\begin{figure}[htbp]
\centering
\makebox[\textwidth][c]{\includegraphics[width=0.9\paperwidth]{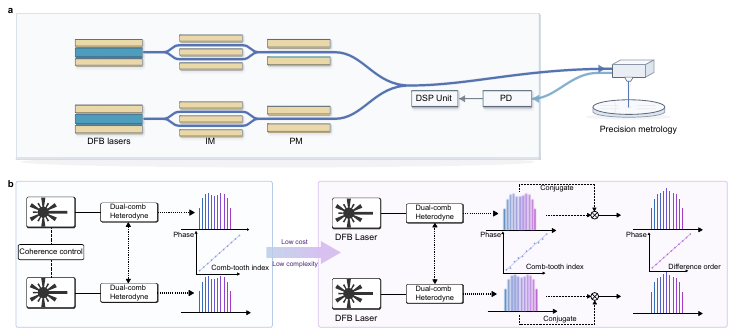}}
\caption{Architecture schematic and operating concept of CPNC-enabled low-cost dual-comb ranging. \textbf{a}, Schematic of the low-cost frequency-domain DCR architecture based on free-running DFB lasers, electro-optic intensity and phase modulation, photodetection and digital processing for precision metrology. DCR, dual-comb ranging; DFB, distributed-feedback; IM, intensity modulator; PM, phase modulator; PD, photodetector; DSP, digital signal processing. \textbf{b}, Conventional frequency-domain DCR (left) uses coherence-controlled sources to maintain resolvable RF-comb teeth and sufficient probe--reference phase correlation for phase-slope retrieval. The proposed architecture (right) uses free-running DFB lasers and applies CPNC to the complete multitone RF signal before complex spectral-component extraction and phase evaluation. The self-conjugate operation removes the common laser phase factor and optical-frequency-offset term while preserving the distance-dependent phase slope in a difference-frequency comb, thereby reducing source-stabilisation and phase-management requirements. CPNC, conjugate phase-noise cancellation; RF, radio-frequency.}\label{fig:system-concept}
\end{figure}

Here we demonstrate a low-cost frequency-domain DCR architecture that combines independent free-running megahertz-linewidth distributed-feedback (DFB) lasers with pre-extraction conjugate phase-noise cancellation (CPNC). By constructing a difference-frequency comb before the complex coefficients and phases of individual RF-comb teeth are extracted, the architecture cancels the common laser phase factor and optical-frequency-offset term while retaining the distance-dependent phase slope. The CPNC branch therefore operates without data from an auxiliary laser-beat channel or explicit common-phase tracking. Using electro-optic combs seeded by DFB lasers with linewidths of $12~\mathrm{MHz}$ and $9~\mathrm{MHz}$, the architecture reaches an Allan deviation of $219~\mathrm{nm}$ at $246~\mu\mathrm{s}$. Measurements across four linewidth configurations and a stepwise displacement experiment further demonstrate robust ranging and displacement recovery. Together, these results show that low-cost free-running sources can support high-precision frequency-domain DCR with reduced source-stabilisation and phase-management requirements. Figure~\ref{fig:system-concept}a schematically illustrates the low-cost DFB-laser-based ranging architecture, and Fig.~\ref{fig:system-concept}b summarises the CPNC operating concept.

\FloatBarrier

\section{Results}\label{sec2}

\subsection{Principle of conjugate phase-noise cancellation}

Frequency-domain DCR infers distance from the phase slope across RF-comb teeth \cite{martin2022performance,wang2025nanometric}. When phase-noise broadening approaches the $30~\mathrm{MHz}$ RF-comb spacing used here, neighbouring teeth partially overlap within the acquisition window. Finite-window extraction then mixes their complex coefficients, converting tooth-common phase fluctuations into comb-index-dependent phase errors. Subsequent probe--reference subtraction cannot generally recover the unmixed phases, so slope fitting and phase unwrapping can produce erroneous distances. Conventional phase subtraction relies on probe--reference phase correlation for common-mode suppression, so reduced correlation can increase the residual phase error. Supplementary Note 1 derives how finite-window phase-noise mixing produces this error and how it projects onto the retrieved distance.

For either the probe or reference channel, $q\in\{p,r\}$, the complex multitone RF-comb signal can be written as
\begin{equation}
\widetilde{i}_q(t)=
e^{j[\Delta\omega t+\eta_q(t)]}
\sum_n B_{q,n}e^{jn\Delta\Omega t}.
\end{equation}
Here, $\eta_q(t)$ is the relative laser phase fluctuation shared by all RF-comb teeth in channel $q$, including the accumulated phase arising from time-dependent variations of the optical frequency offset. The quantity $\Delta\omega$ denotes the time-independent component of that offset, whereas $\Delta\Omega$ is the repetition angular-frequency difference. The ideal coefficient $B_{q,n}$ contains the optical amplitudes, static system response and distance-dependent phase across the comb-tooth index. Crucially, CPNC operates on the complete multitone signal before the complex coefficients and phases of the individual RF-comb teeth are extracted.

CPNC forms the pointwise self-conjugate product of the complete multitone signal:
\begin{equation}
\begin{aligned}
y_q(t)
&=\widetilde{i}_q(t)\widetilde{i}_q^*(t)\\
&=e^{j[\Delta\omega t+\eta_q(t)]}
e^{-j[\Delta\omega t+\eta_q(t)]}
\sum_n\sum_k B_{q,n}B_{q,k}^*
e^{j(n-k)\Delta\Omega t}\\
&=\sum_n\sum_k B_{q,n}B_{q,k}^*
e^{j(n-k)\Delta\Omega t}.
\end{aligned}
\end{equation}
The complete tooth-common phase factor therefore cancels identically before complex spectral-component extraction and phase evaluation. Because the cancellation holds independently of the temporal form of $\eta_q(t)$, it also removes the common phase evolution associated with RF-comb centre-frequency drift without requiring its instantaneous value. Because CPNC cancels $\eta_q(t)$ within each channel before that subtraction, it does not require $\eta_p(t)$ and $\eta_r(t)$ to remain correlated when the path imbalance exceeds the coherence scale. Cross terms with $n\neq k$ remain and form a difference-frequency comb indexed by the difference order $m=n-k$. For a pair of teeth separated by $m$, the distance-dependent part of the product can be written as
\begin{equation}
B_{q,k+m}B_{q,k}^* \propto e^{j2\pi m f_{r,\mathrm{sig}}\tau_q},
\end{equation}
up to static amplitude and instrumental phase factors. Thus, after grouping all cross terms with the same $m$, the $m$th difference-frequency component carries a phase that remains linear in $m$. A full derivation is provided in Supplementary Note 1.

Applying CPNC separately to the probe and reference channels yields two difference-frequency combs. Their phase difference at order $m$ contains a time-invariant instrumental contribution and a distance-dependent term,
\begin{equation}
\Phi_m\approx \phi_m^{\mathrm{inst}}+2\pi m f_{r,\mathrm{sig}}\Delta\tau,
\qquad
d_{\mathrm{retrieved}}=\frac{c\Delta\tau}{2}+d_0,
\end{equation}
where $\Delta\tau=\tau_p-\tau_r$ is the relative delay, $c$ is the speed of light and $d_0$ is a fixed distance offset associated with the static differential instrumental phase response $\phi_m^{\mathrm{inst}}$. Here, $f_{r,\mathrm{sig}}=25.000~\mathrm{GHz}$ is the signal-comb repetition rate. A linear fit of $\Phi_m$ against $m$ therefore tracks the delay up to this fixed offset. The fixed offset $d_0$ does not affect the Allan deviation, single-frame fluctuation or relative-displacement metrics reported here. The distance-dependent phase slope is governed by $f_{r,\mathrm{sig}}$, whereas $\Delta f_r$ sets only the radio-frequency spacing of the difference-frequency components. Details of complex spectral-component extraction, phase-slope fitting and performance-metric evaluation are provided in Supplementary Note 2.

The following experiments test phase-slope recovery with megahertz-linewidth DFB lasers.

\subsection{High-precision ranging with MHz-linewidth lasers}

\begin{figure}[!htbp]
\centering
\makebox[\textwidth][c]{\includegraphics[width=0.87\paperwidth]{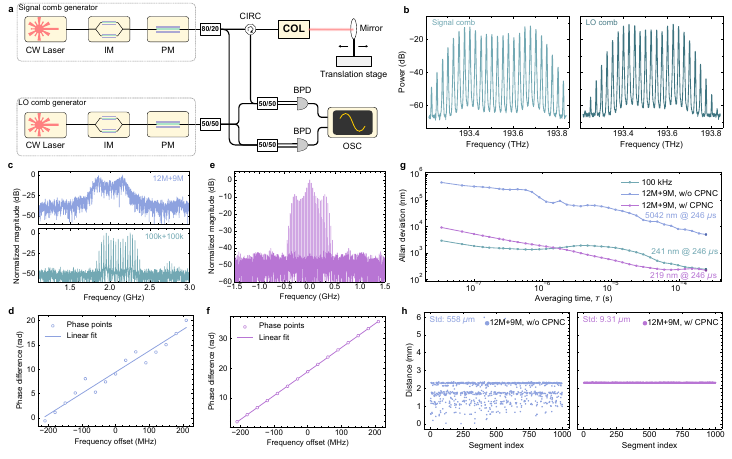}}
\caption{CPNC-enabled high-precision dual-comb ranging with MHz-linewidth lasers. \textbf{a}, Experimental configuration. CW, continuous-wave laser; IM, intensity modulator; PM, phase modulator; CIRC, optical circulator; COL, collimator; BPD, balanced photodetector; OSC, oscilloscope; EO, electro-optic; LO, local oscillator. The dashed boxes indicate the signal and LO EO-comb generators; coupler labels give the splitting ratios. Microwave drivers, the auxiliary laser-beat channel and the alignment lens are omitted. \textbf{b}, Measured optical spectra of the signal and LO combs. \textbf{c}, Stacked RF-comb spectra for the $12~\mathrm{MHz}+9~\mathrm{MHz}$ configuration (upper) and the $100~\mathrm{kHz}+100~\mathrm{kHz}$ reference (lower). The upper spectrum contains broadened and partially overlapping teeth, whereas the lower spectrum shows resolved teeth. \textbf{d}, Raw phase-slope fit for the $12~\mathrm{MHz}+9~\mathrm{MHz}$ measurement after tooth-wise complex-coefficient extraction. \textbf{e}, Difference-frequency comb obtained after CPNC. \textbf{f}, CPNC phase-slope fit using 15 valid difference-frequency components. \textbf{g}, Allan deviation for the narrow-linewidth reference and the $12~\mathrm{MHz}+9~\mathrm{MHz}$ measurement before and after CPNC. At $246~\mu\mathrm{s}$, CPNC reduces the Allan deviation from $5042~\mathrm{nm}$ to $219~\mathrm{nm}$; the reference gives $241~\mathrm{nm}$. \textbf{h}, First 1,000 static-distance estimates before and after CPNC. Single-frame standard deviations calculated from all 153,903 frames decrease from $558~\mu\mathrm{m}$ to $9.31~\mu\mathrm{m}$. CPNC, conjugate phase-noise cancellation; DCR, dual-comb ranging; RF, radio-frequency; w/o, without CPNC; w/, with CPNC.}\label{fig:dcr-overview}
\end{figure}

To validate the CPNC-enabled architecture, we performed a static-ranging experiment using electro-optic combs generated from independent free-running DFB seed lasers (Fig.~\ref{fig:dcr-overview}a). The two combs were driven at repetition rates of $25.000~\mathrm{GHz}$ and $25.030~\mathrm{GHz}$, giving $\Delta f_r=30~\mathrm{MHz}$; their optical spectra are shown in Fig.~\ref{fig:dcr-overview}b. The signal comb was divided into a free-space probe path reflected from a static mirror and a fixed reference path. The probe path contained approximately $6~\mathrm{m}$ more fiber than the reference path and a $0.625~\mathrm{m}$ one-way free-space path. Including the free-space round trip and using a fiber group index of 1.468 gives an estimated signal-path differential delay of $33.6~\mathrm{ns}$.

\Needspace{3\baselineskip}
The returned probe and reference combs were separately heterodyned with the local-oscillator (LO) comb and recorded, generating RF dual-comb signals for phase-slope ranging. Because the two seed lasers were free-running and not carrier-frequency locked, their optical frequency offset was set near $2~\mathrm{GHz}$ but allowed to drift during acquisition. This optical frequency offset sets the centre frequency of the original RF-comb, denoted by $f_c$ after mapping to the RF domain. An auxiliary laser-beat channel tracked this RF-comb centre frequency only to localise RF-comb teeth in the raw-processing branch used for comparison; its data were not used in the CPNC branch, and its configuration is shown in Supplementary Fig. 1 and detailed in Supplementary Note 3.

To assess the effect of source linewidth, we compared the $12~\mathrm{MHz}+9~\mathrm{MHz}$ configuration with a $100~\mathrm{kHz}+100~\mathrm{kHz}$ narrow-linewidth reference. In the notation $X+Y$, $X$ and $Y$ denote the signal-comb and LO-comb seed-laser linewidths, respectively. Delayed self-heterodyne interferometry (DSHI) measurements of the three MHz-linewidth DFB seed lasers are shown in Supplementary Fig. 2 and detailed in Supplementary Note 4.

\Needspace{5\baselineskip}
Assuming a Lorentzian line shape, the $12~\mathrm{MHz}$ signal-laser linewidth corresponds to an estimated $1/e$ field-correlation time of approximately $26.5~\mathrm{ns}$. The estimated signal-path differential delay therefore exceeds this coherence timescale, indicating substantially reduced probe--reference phase correlation. Figure~\ref{fig:dcr-overview}c presents the corresponding RF-comb spectra in a stacked format. The narrow-linewidth reference in the lower panel exhibits individually resolved RF-comb teeth, whereas the MHz-linewidth configuration in the upper panel produces broadened and partially overlapping teeth. Under this condition, conventional tooth-wise complex-coefficient extraction mixes contributions from neighbouring RF-comb teeth within the finite analysis window, degrading the phases subsequently obtained from those coefficients and the resulting phase-slope fit. This comparison therefore provides a stringent test of CPNC.

The resulting phase-slope degradation is evident in the raw-processing result for the $12~\mathrm{MHz}+9~\mathrm{MHz}$ configuration (Fig.~\ref{fig:dcr-overview}d). The extracted comb-tooth phases deviate substantially from the linear fit, making the phase-slope estimate unreliable. Applying CPNC to the same recorded signals produces a resolved difference-frequency comb without the drifting optical-frequency-offset term (Fig.~\ref{fig:dcr-overview}e). The phases of 15 valid difference-frequency components then exhibit a clear linear dependence on the difference order (Fig.~\ref{fig:dcr-overview}f), demonstrating retention of the distance-dependent phase slope. The influence of fitted component number and spectral span is examined experimentally in Supplementary Fig. 3 and Supplementary Note 5, which also provides an idealized reference scaling and states its assumptions. These results show that CPNC recovers the phase observable required for frequency-domain DCR when the original RF-comb teeth are broadened and partially overlapping.

The recovered phase slope translates into improved static ranging precision. Figure~\ref{fig:dcr-overview}g compares the Allan deviations of the raw and CPNC-processed $12~\mathrm{MHz}+9~\mathrm{MHz}$ measurements with the $100~\mathrm{kHz}+100~\mathrm{kHz}$ narrow-linewidth reference. At an averaging time of $246~\mu\mathrm{s}$, the Allan deviation decreases from $5042~\mathrm{nm}$ without CPNC to $219~\mathrm{nm}$ after CPNC, corresponding to an approximately 23-fold improvement. The narrow-linewidth reference gives $241~\mathrm{nm}$ at the same averaging time. CPNC therefore recovers submicrometre short-term stability with MHz-linewidth DFB lasers, yielding an Allan deviation comparable to the narrow-linewidth reference at $246~\mu\mathrm{s}$.

The improvement was also evident at the individual-frame level (Fig.~\ref{fig:dcr-overview}h). The first 1,000 consecutive distance estimates are shown for visual clarity, whereas the single-frame statistics were calculated using all 153,903 segmented frames. The raw $12~\mathrm{MHz}+9~\mathrm{MHz}$ measurement had a standard deviation of $558~\mu\mathrm{m}$, which decreased to $9.31~\mu\mathrm{m}$ after CPNC, corresponding to an approximately 60-fold reduction in frame-to-frame distance fluctuation. This comparison complements the Allan deviation analysis by showing that CPNC improves the consistency of individual-frame distance retrieval before temporal averaging.

\FloatBarrier

\subsection{Precise stepwise displacement tracking with MHz-linewidth lasers}

To demonstrate controlled stepwise displacement tracking using the CPNC-recovered phase slope, we translated a plane mirror mounted on a differential-micrometer flexure stage in $100~\mu\mathrm{m}$ increments over a total range of $1~\mathrm{mm}$. At each position, distance estimates acquired over $1~\mu\mathrm{s}$ were averaged to obtain the plotted value, and all positions were approached from the same direction to minimize mechanical hysteresis.

\begin{figure}[htbp]
\centering
\makebox[\textwidth][c]{\includegraphics[width=0.9\paperwidth]{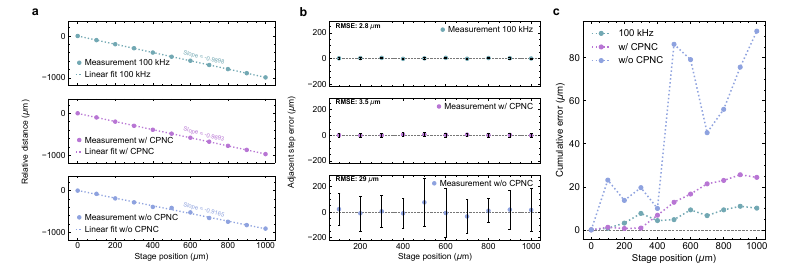}}
\caption{Stepwise displacement tracking with MHz-linewidth lasers. \textbf{a}, Measured relative distance as a function of translation-stage position for the $100~\mathrm{kHz}+100~\mathrm{kHz}$ narrow-linewidth reference, the CPNC-processed $12~\mathrm{MHz}+9~\mathrm{MHz}$ measurement and the raw $12~\mathrm{MHz}+9~\mathrm{MHz}$ measurement. The plane mirror was translated in $100~\mu\mathrm{m}$ steps over a total travel range of $1~\mathrm{mm}$. At each position, the plotted value was obtained by averaging $1~\mu\mathrm{s}$ of data. Solid markers denote measured distances and dashed lines denote linear fits. Because the stage motion decreases the optical path length, the ideal slope is $-1$. The fitted slopes are $-0.9898$, $-0.9693$ and $-0.9165$, respectively. \textbf{b}, Adjacent-step error for the three measurements. The adjacent-step error is defined as the difference between the DCR-measured displacement increment between two neighbouring stage positions and the corresponding $100~\mu\mathrm{m}$ stage increment. The adjacent-step RMSEs are $2.8~\mu\mathrm{m}$, $3.5~\mu\mathrm{m}$ and $29~\mu\mathrm{m}$, respectively. Error bars denote the propagated standard deviation from the two neighbouring stage positions. \textbf{c}, Signed cumulative displacement error relative to the first stage position. The maximum errors are $11.0~\mu\mathrm{m}$ for the narrow-linewidth reference, $25.6~\mu\mathrm{m}$ after CPNC and $92.3~\mu\mathrm{m}$ without CPNC. CPNC, conjugate phase-noise cancellation; DCR, dual-comb ranging; RMSE, root-mean-square error.}\label{fig:dcr-results}
\end{figure}

Figure~\ref{fig:dcr-results}a shows the measured relative distance as a function of stage position. Because the stage translation shortened the optical path length, the ideal displacement response had a slope of $-1$. All three measurements exhibited high linearity, with $R^2$ values ranging from $0.9959$ to $1.0000$. Their fitted slopes, however, differed in their agreement with the ideal response. The narrow-linewidth reference yielded a slope of $-0.9898$, corresponding to a scale-factor deviation of $1.02\%$. The CPNC-processed $12~\mathrm{MHz}+9~\mathrm{MHz}$ measurement yielded a slope of $-0.9693$, whereas the raw measurement yielded $-0.9165$. These slopes correspond to scale-factor deviations of $3.07\%$ and $8.35\%$, respectively. CPNC therefore reduced the scale-factor deviation under MHz-linewidth operation and brought the displacement response closer to the narrow-linewidth reference.

To quantify local step-recovery performance, we calculated the adjacent-step error as the difference between each DCR-measured displacement increment and the corresponding $100~\mu\mathrm{m}$ stage increment. This metric evaluates each step independently of the displacement relative to the initial stage position.

\Needspace{6\baselineskip}
As shown in Fig.~\ref{fig:dcr-results}b, the adjacent-step root-mean-square error (RMSE) was $2.8~\mu\mathrm{m}$ for the $100~\mathrm{kHz}+100~\mathrm{kHz}$ reference and $3.5~\mu\mathrm{m}$ for the CPNC-processed $12~\mathrm{MHz}+9~\mathrm{MHz}$ measurement. The raw measurement yielded a substantially larger RMSE of $29~\mu\mathrm{m}$. CPNC therefore reduced the adjacent-step RMSE by approximately 8.3-fold and recovered the $100~\mu\mathrm{m}$ increments with micrometre-scale performance close to the narrow-linewidth reference.

To examine the displacement response over the full scan range, we calculated the signed displacement error relative to the initial stage position (Fig.~\ref{fig:dcr-results}c). This quantity compares the DCR-measured displacement from the starting point with the corresponding stage displacement. It therefore captures scale-factor and position-dependent deviations over increasing displacement spans, rather than the error of an individual step. The maximum signed cumulative displacement error was $92.3~\mu\mathrm{m}$ for the raw $12~\mathrm{MHz}+9~\mathrm{MHz}$ measurement, decreasing to $25.6~\mu\mathrm{m}$ after CPNC, compared with $11.0~\mu\mathrm{m}$ for the narrow-linewidth reference. Together, the local and full-range analyses demonstrate precise recovery of unidirectional stepwise displacements under MHz-linewidth operation.

\FloatBarrier

\subsection{Ranging stability across laser-linewidth configurations}

\begin{figure}[htbp]
\centering
\makebox[\textwidth][c]{\includegraphics[width=0.9\paperwidth]{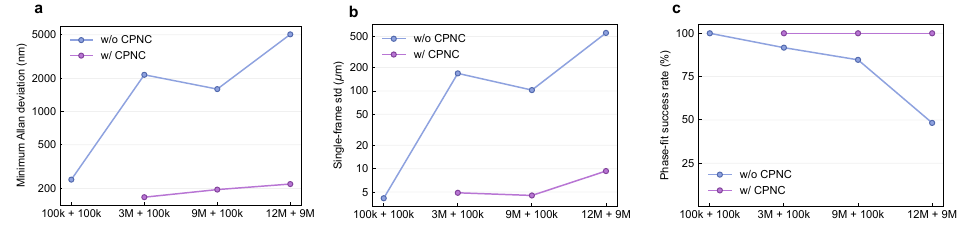}}
\caption{CPNC performance across laser-linewidth configurations. \textbf{a}, Minimum Allan deviations for the $100~\mathrm{kHz}+100~\mathrm{kHz}$ narrow-linewidth reference without CPNC and for three megahertz-linewidth configurations with and without CPNC. In each configuration, the first and second values denote the signal-comb and LO-comb seed-laser linewidths, respectively. The minima occur at configuration-dependent averaging times of $37.7$--$246~\mu\mathrm{s}$. \textbf{b}, Single-frame distance standard deviations for the same configurations. CPNC reduces the values for the three megahertz-linewidth configurations to $4.52$--$9.31~\mu\mathrm{m}$, compared with $4.16~\mu\mathrm{m}$ for the narrow-linewidth reference. The vertical axes in \textbf{a} and \textbf{b} are logarithmic. \textbf{c}, Phase-fit success rate for distance retrieval. A frame is counted as successful when the phase-slope-fit RMSE is below $1~\mathrm{rad}$, and the phase-fit success rate is calculated as $N_{\mathrm{success}}/N_{\mathrm{total}}\times100\%$. CPNC yields a phase-fit success rate of $100\%$ for all three megahertz-linewidth configurations. Blue markers denote measurements without CPNC, and purple markers denote CPNC-processed measurements. CPNC, conjugate phase-noise cancellation; LO, local oscillator; RMSE, root-mean-square error.}\label{fig:dcr-fig4}
\end{figure}

To determine whether CPNC maintained ranging performance across different source linewidths, we evaluated four linewidth configurations (Fig.~\ref{fig:dcr-fig4}a): the $100~\mathrm{kHz}+100~\mathrm{kHz}$ narrow-linewidth reference, two mixed-linewidth configurations and the $12~\mathrm{MHz}+9~\mathrm{MHz}$ configuration. Without CPNC, the minimum Allan deviations for the three megahertz-linewidth configurations ranged from $1.60$ to $5.04~\mu\mathrm{m}$, compared with $241~\mathrm{nm}$ for the narrow-linewidth reference. With CPNC, the corresponding minima decreased to $167$--$219~\mathrm{nm}$ at configuration-dependent averaging times of $37.7$--$246~\mu\mathrm{s}$. The $12~\mathrm{MHz}+9~\mathrm{MHz}$ configuration reached $219~\mathrm{nm}$ at $246~\mu\mathrm{s}$. Thus, all three CPNC-processed megahertz-linewidth configurations achieved sub-$250~\mathrm{nm}$ minimum Allan deviations, including operation with two independent megahertz-linewidth DFB lasers. Detailed Allan-deviation curves and single-frame distance sequences for the two mixed-linewidth configurations are provided in Supplementary Figs. 4--7 and Supplementary Note 6.

We next compared the single-frame distance standard deviations across the linewidth configurations (Fig.~\ref{fig:dcr-fig4}b). Without CPNC, the standard deviations for the $3~\mathrm{MHz}+100~\mathrm{kHz}$, $9~\mathrm{MHz}+100~\mathrm{kHz}$ and $12~\mathrm{MHz}+9~\mathrm{MHz}$ configurations were $168$, $103$ and $558~\mu\mathrm{m}$, respectively.

\Needspace{4\baselineskip}
After CPNC, the corresponding values decreased to $4.89$, $4.52$ and $9.31~\mu\mathrm{m}$. The CPNC-processed results therefore remained below $10~\mu\mathrm{m}$ across all three megahertz-linewidth configurations, compared with $4.16~\mu\mathrm{m}$ for the narrow-linewidth reference. This frame-level comparison shows that the improvement did not depend on temporal averaging across successive frames.

Finally, we assessed frame-wise phase-slope fitting using the phase-fit success rate (Fig.~\ref{fig:dcr-fig4}c). A frame was counted as successful when the phase-slope-fit RMSE was below $1~\mathrm{rad}$. Without CPNC, the phase-fit success rate decreased across the megahertz-linewidth configurations and fell below $50\%$ for the $12~\mathrm{MHz}+9~\mathrm{MHz}$ configuration. After CPNC, every analysed frame in all three megahertz-linewidth configurations satisfied the fitting criterion, giving a phase-fit success rate of $100\%$. The narrow-linewidth reference without CPNC also achieved a phase-fit success rate of $100\%$. Together with the Allan deviation and single-frame statistics, these results show that CPNC maintained both phase-fit reliability and short-term ranging stability across the tested linewidth configurations.

\FloatBarrier

\section{Discussion}\label{sec:discussion}

In this work, we experimentally demonstrated a CPNC-enabled low-cost frequency-domain dual-comb ranging architecture based on independent free-running megahertz-linewidth DFB lasers. CPNC forms a self-conjugate product of each multitone RF signal, converting absolute comb-tooth phases into inter-tooth differential phases. This transformation cancels the common laser phase factor and drifting optical-frequency-offset term before tooth-wise complex-coefficient and phase extraction while preserving the distance-dependent phase slope. With seed-laser linewidths of $12~\mathrm{MHz}$ and $9~\mathrm{MHz}$, the system reached an Allan deviation of $219~\mathrm{nm}$ at $246~\mu\mathrm{s}$ and reduced the single-frame distance standard deviation from $558$ to $9.31~\mu\mathrm{m}$. These results demonstrate that CPNC suppresses common phase fluctuations without data from the auxiliary laser-beat channel, thereby reducing reliance on highly coherent sources, probe--reference phase correlation and separate phase-tracking procedures.

Representative implementations include stabilised fibre-comb ranging \cite{coddington2009rapid}, soliton-microcomb ranging \cite{trocha2018ultrafast,wang2025nanometric}, dispersive-Fourier-transform retrieval \cite{chang2024dispersive} and frequency-locked dual-microcomb ranging \cite{yan2025rapid}. Specifically, as shown in Table~\ref{tab:dcr-comparison}, our architecture achieves frequency-domain phase-slope ranging with independent free-running DFB seed lasers having linewidths of $12~\mathrm{MHz}$ and $9~\mathrm{MHz}$, without active source stabilisation. The CPNC branch additionally operates without monitor-derived RF-comb centre-frequency tracking. A previous free-running implementation maintained RF-comb-tooth resolvability and probe--reference phase correlation through large RF-comb spacing and a probe--reference path imbalance within the coherence length \cite{trocha2022ultrafast}. CPNC instead removes the common laser phase factor and optical-frequency-offset term before extracting the complex coefficients and phases of individual RF-comb teeth.

\Needspace{4\baselineskip}
This combination enables submicrometre ranging in frequency-domain DCR using independent free-running megahertz-linewidth DFB sources, without requiring fully resolved original RF-comb teeth or the probe--reference path imbalance to remain within the coherence length.

\begin{table}[htbp]
\caption{Source-coherence and phase-management strategies in representative dual-comb ranging systems.}
\label{tab:dcr-comparison}
\footnotesize
\setlength{\tabcolsep}{3pt}
\renewcommand{\arraystretch}{1.10}
\begin{tabularx}{\textwidth}{>{\raggedright\arraybackslash}p{0.10\textwidth}>{\raggedright\arraybackslash}p{0.19\textwidth}>{\raggedright\arraybackslash}p{0.31\textwidth}>{\raggedright\arraybackslash}X}
\toprule
Year & Comb source & Source/coherence control & Phase/offset handling \\
\midrule
2018\cite{trocha2018ultrafast} & DKS microcombs & Independent; free-running pumps & Phase slope + DSP \\
2022\cite{trocha2022ultrafast} & QD-MLLDs & Independent; free-running ($\sim15~\mathrm{MHz}$) & Resolved RF teeth; correlated phase subtraction\textsuperscript{a} \\
2024\cite{chang2024dispersive} & Fibre combs & RF phase-locked & DFT interferometry \\
2025\cite{wang2025nanometric} & CP solitons & VFL; passive mutual coherence & Phase slope; direct Fourier transform \\
2025\cite{yan2025rapid} & Dual microcombs & Common $^{87}$Rb-locked CW pump; RF-locked repetition rates & AOM-set pump offset; phase-slope fitting \\
This work & DFB-seeded EO combs & Independent; free-running ($12~\mathrm{MHz}+9~\mathrm{MHz}$) & Pre-extraction CPNC; no optical-frequency-offset tracking \\
\bottomrule
\end{tabularx}
\vspace{2pt}
\parbox{\textwidth}{\scriptsize \textsuperscript{a} Phase noise remained correlated because the probe--reference path imbalance was shorter than the coherence length. DKS, dissipative Kerr soliton; QD-MLLD, quantum-dash mode-locked laser diode; CP, counter-propagating; VFL, Vernier frequency locking; DFT, dispersive Fourier transform; DSP, digital signal processing; AOM, acousto-optic modulator; CW, continuous-wave; EO, electro-optic; RF, radio-frequency.}
\end{table}

Across the three megahertz-linewidth configurations in Fig.~\ref{fig:dcr-fig4}, CPNC yielded $100\%$ phase-fit success and sub-$250$-nm minimum Allan deviations, whereas the remaining configuration-dependent differences in single-frame precision are consistent with additive receiver noise and non-common component fluctuations that remain after CPNC and project onto the fitted phase slope (Supplementary Note 1). Broader and flatter electro-optic combs with higher received signal-to-noise ratio should improve component-phase precision \cite{martin2022performance}, while a wider usable fitting span should increase phase-slope leverage \cite{wang2025nanometric}.

The combination of low-cost free-running DFB lasers and pre-extraction CPNC establishes a frequency-domain DCR architecture with reduced source-stabilisation and phase-management complexity. Further integration of DFB seed lasers \cite{shamsansari2022electrically} with electro-optic comb generators \cite{zhang2019broadband,hu2022highefficiency} and photonic routing components \cite{liu2025design,shen2018reconfigurable}, together with real-time CPNC processing, could support compact, low-complexity implementations and dynamic ranging. More broadly, the same differential-phase architecture may be extended to high-speed vibrometry and displacement sensing \cite{teleanu2017electro}, as well as integrated dual-comb spectroscopy \cite{dutt2018onchip,chang2022integrated}.

\section{Methods}\label{sec:methods}

\subsection{Detailed experimental setup for dual-comb ranging}\label{sec:methods-setup}

The MHz-linewidth sources were free-running DFB lasers operated without optical phase or carrier-frequency locking, whereas two $100~\mathrm{kHz}$ external-cavity lasers (IQS-636, EXFO) were used in the narrow-linewidth comparison configurations.
The experimental setup comprised two independent free-running lasers, two electro-optic comb generators, a free-space ranging arm, two balanced photodetectors, an auxiliary single photodetector and a high-speed oscilloscope. Two continuous-wave seed lasers provided the optical inputs, with linewidths selected according to each tested configuration. Throughout this work, a linewidth configuration of $X+Y$ denotes the signal-comb and LO-comb seed-laser linewidths, respectively. In the mixed-linewidth configurations, the broader-linewidth laser seeded the signal comb and the $100~\mathrm{kHz}$ external-cavity laser seeded the LO comb. Thus, the $3~\mathrm{MHz}+100~\mathrm{kHz}$ and $9~\mathrm{MHz}+100~\mathrm{kHz}$ configurations used MHz-linewidth signal combs heterodyned with a narrow-linewidth LO comb. In the $12~\mathrm{MHz}+9~\mathrm{MHz}$ configuration, the $12~\mathrm{MHz}$ laser seeded the signal comb and the $9~\mathrm{MHz}$ laser seeded the LO comb. Each seed laser was sent to an electro-optic comb generator consisting of cascaded lithium-niobate intensity and phase modulators. The modulators were commercial EOspace devices, including AX-0MVS-40 intensity modulators and PM-5VEK-40 phase modulators. The two comb generators were driven at repetition frequencies of $25.000~\mathrm{GHz}$ and $25.030~\mathrm{GHz}$, respectively, producing two electro-optic combs with a repetition-rate difference of $30~\mathrm{MHz}$. Polarisation controllers were used before the modulators and before heterodyne detection to optimise the modulation state and beating efficiency. An auxiliary laser-beat channel, implemented with a single photodetector (XPDV3120, Finisar), recorded the instantaneous optical frequency offset between the two seed lasers. Further optical-routing details are provided in Supplementary Note 3. This signal was used solely in the raw-processing branch and is not required for CPNC processing or CPNC-based distance retrieval.

\Needspace{12\baselineskip}
The signal comb was split into a probe path and a reference path. In the probe path, the optical beam was coupled into free space using a fiber collimator (F810-APC, Thorlabs), passed through an achromatic doublet lens (AC254-500-C-ML, Thorlabs) to facilitate beam alignment and efficient recoupling, reflected by a plane mirror located approximately $0.625~\mathrm{m}$ from the collimator and coupled back into fiber. The probe path contained approximately $6~\mathrm{m}$ more fiber than the reference path. Including the free-space round trip and using a fiber group index of 1.468 gives an estimated signal-path differential delay of $33.6~\mathrm{ns}$. For displacement measurements, the plane mirror was mounted on a manual flexure translation stage with a differential micrometer (NFL5D/M, Thorlabs). The stage was advanced in one direction during the stepwise measurements to reduce hysteresis. The reference path was kept fixed and was used to generate the reference dual-comb signal. The returned probe comb and the reference comb were separately combined with the LO comb and detected by two 10-GHz balanced photodetectors (KG-BPD-10G-A, Beijing Kangguan Century Optoelectronics Technology Co., Ltd.). Across all linewidth configurations, the total optical powers of the returned probe comb and the LO comb, measured immediately before optical combination and balanced photodetection, were maintained at approximately $-0.5~\mathrm{dBm}$ and $-3.4~\mathrm{dBm}$, respectively. The electrical outputs of the photodetectors were recorded by a high-speed oscilloscope (DSAX96204Q, Keysight Technologies) operated at a sampling rate of $10~\mathrm{GSa/s}$. The acquired time-domain RF dual-comb signals were stored and processed offline for phase extraction, conjugate phase-noise cancellation and distance retrieval.

\subsection{Digital signal processing}\label{sec:methods-dsp}

The recorded oscilloscope traces were processed using a segment-wise digital signal-processing procedure. Each measurement contained three electrical channels: the probe channel, the reference channel and an auxiliary laser-beat channel. The probe and reference channels provided the ranging signals. The auxiliary laser-beat channel was used only in the raw-processing branch to track the instantaneous RF-comb centre frequency associated with drift of the optical frequency offset between the two free-running lasers. The recorded probe and reference traces were first cropped to the selected temporal window and divided into processing segments with a duration of $T_{\mathrm{seg}}=1/\Delta f_r$, where $\Delta f_r=30~\mathrm{MHz}$. For RF-comb centre-frequency tracking in the raw-processing branch, ten consecutive processing segments were grouped into one tracking window. In each tracking window, the RF-comb centre frequency was estimated from the dominant peak in the Fourier spectrum of the auxiliary laser-beat channel and assigned to the corresponding probe and reference segments.

\Needspace{7\baselineskip}
Two processing branches were then applied to the same recorded probe and reference traces. In the raw-processing branch, the complex coefficients of the selected RF-comb teeth were obtained by direct Fourier evaluation at their target frequencies using the monitor-derived RF-comb centre-frequency estimate for tooth localisation. The probe--reference phase difference was calculated for each comb tooth, unwrapped along the comb-tooth index and fitted with a linear function to retrieve the distance-dependent phase slope.

\Needspace{8\baselineskip}
In the CPNC branch, the probe and reference traces were first converted into analytic signals using a Hilbert transform and then multiplied pointwise by their complex conjugates to generate difference-frequency combs. No data from the auxiliary laser-beat channel or monitor-derived RF-comb centre-frequency estimate were used in this branch. The selected difference-frequency components were then extracted using the same direct Fourier evaluation, and the probe--reference phase difference was calculated, unwrapped and linearly fitted as a function of the difference order.

\Needspace{5\baselineskip}
The reported stability and relative-displacement metrics were evaluated without independent instrumental-phase calibration because the time-invariant differential response of the probe and reference channels contributes only a fixed distance offset. The retrieved distance sequences from the raw-processing and CPNC branches were used for performance evaluation, including distance fluctuation, phase-fit success rate and Allan deviation.

\backmatter

\bmhead{Data availability}

The data that support the findings of this study are available from the corresponding author upon reasonable request.

\bmhead{Code availability}

The code used for data processing is available from the corresponding author upon reasonable request.

\bmhead{Acknowledgements}

This work was supported by the National Natural Science Foundation of China under Grant No. 62271010 and by the High-performance Computing Platform of Peking University.

\Needspace{10\baselineskip}
\bmhead{Author contributions}

Y.Y. conceived the preliminary idea of conjugate phase-noise cancellation. Y.Y. and Z.W. jointly performed the DCR experiments and analysed the data. Y.Zou, C.Z. and X.X. constructed the electro-optic comb generators. X.C. and D.L. performed the laser-linewidth measurements. Y.Y., Z.W., S.C., S.P. and F.Z. discussed the experimental architecture. Y.Y., Z.W., Y.Zhu, X.Z. and F.Z. constructed the basic concept of dual-comb ranging with megahertz-linewidth lasers and discussed prospective development trends. Y.Y., Z.W. and F.Z. prepared the original draft. F.Z. supervised the project. All authors reviewed and revised the manuscript and approved the final version. Y.Y. and Z.W. contributed equally to this work.

\bmhead{Competing interests}

Y.Y., Z.W. and F.Z. have filed a patent application on the dual-comb-based lidar signal generation and detection method and system: CN202610175865.4, filed 6 February 2026. The remaining authors declare no competing interests.

\bmhead{Additional information}

Supplementary information will be available for this paper. Correspondence and requests for materials should be addressed to F.Z.

\bibliography{sn-bibliography}

\clearpage
% Supplementary Information appended to main.tex.
\setcounter{figure}{0}
\setcounter{table}{0}
\setcounter{equation}{0}
\renewcommand{\thefigure}{S\arabic{figure}}
\renewcommand{\figurename}{Fig.}
\renewcommand{\tablename}{Supplementary table}
\renewcommand{\thetable}{\arabic{table}}
\renewcommand{\theHfigure}{S\arabic{figure}}
\renewcommand{\theHtable}{S\arabic{table}}
\renewcommand{\theHequation}{S\arabic{equation}}
\captionsetup{font=small,labelfont=bf,labelsep=period}

\thispagestyle{empty}
\begingroup
\centering
\vspace*{0.1cm}
{\normalsize Supplementary Information for\par}
\vspace{0.45cm}
{\Large\bfseries Conjugate phase-noise cancellation enables submicrometre dual-comb ranging with free-running megahertz-linewidth lasers\par}
\vspace{0.8cm}
{\large Yue You$^{1,\dagger}$, Zaifan Wu$^{1,\dagger}$, Yi Zou$^{1}$, Xiang Cai$^{1}$, Yixiao Zhu$^{3}$, Chenbo Zhang$^{1}$, Dan Lu$^{2}$, Siming Chen$^{5,6}$, Shujie Pan$^{7}$, Xian Zhou$^{4}$, Xiaopeng Xie$^{1}$ \& Fan Zhang$^{1,8,*}$\par}
\vspace{0.45cm}
{\normalsize
$^{1}$State Key Laboratory of Photonics and Communications, School of Electronics,\par
Peking University, Beijing 100871, China.\par
$^{2}$State Key Laboratory of Optoelectronic Materials and Devices, Institute of Semiconductors,\par
Chinese Academy of Sciences, Beijing 100083, China.\par
$^{3}$State Key Laboratory of Photonics and Communications, Department of Electronic Engineering,\par
Shanghai Jiao Tong University, Shanghai 200240, China.\par
$^{4}$University of Science and Technology Beijing, Beijing 100083, China.\par
$^{5}$Laboratory of Solid State Optoelectronics Information Technology, Institute of Semiconductors,\par
Chinese Academy of Sciences, Beijing 100083, China.\par
$^{6}$College of Materials Science and Opto-Electronic Technology,\par
University of Chinese Academy of Sciences, Beijing 101804, China.\par
$^{7}$HS Photonics Co., Ltd., Xiangjiang Science \& Technology Innovation Base,\par
Changsha, Hunan 413000, China.\par
$^{8}$Peng Cheng Laboratory, Shenzhen 518055, China.\par}
\vspace{0.75cm}
{\normalsize $^{*}$Corresponding author(s). E-mail(s): \href{mailto:fzhang@pku.edu.cn}{fzhang@pku.edu.cn};\par}
\vspace{0.15cm}
{\normalsize $^{\dagger}$These authors contributed equally to this work.\par}
\vfill
\par
\endgroup
\clearpage

\pagestyle{silayout}
\thispagestyle{silayout}

{\large\bfseries Contents\par}
\vspace{0.7em}
\noindent
\begin{tabularx}{\textwidth}{@{}Xr@{}}
Supplementary Note 1: Propagation of phase noise to distance error and conjugate phase-noise cancellation & \pageref{suppnote:1}\\[0.35em]
Supplementary Note 2: Complex spectral-component extraction, phase-slope retrieval and performance metrics & \pageref{suppnote:2}\\[0.35em]
Supplementary Note 3: Auxiliary laser-beat channel & \pageref{suppnote:3}\\[0.35em]
Supplementary Note 4: Seed-laser linewidth characterization & \pageref{suppnote:4}\\[0.35em]
Supplementary Note 5: Influence of fitted component number and spectral span & \pageref{suppnote:5}\\[0.35em]
Supplementary Note 6: Detailed ranging results for mixed-linewidth laser configurations & \pageref{suppnote:6}\\[0.35em]
Supplementary References & \pageref{supprefs}
\end{tabularx}
\clearpage

\suppnote{1}{Propagation of phase noise to distance error and conjugate phase-noise cancellation}

We consider a dual-comb ranging system in which $\Delta\omega$ denotes the time-independent component of the angular optical-frequency offset between the two optical combs, and $\Delta\Omega$ is their repetition angular-frequency difference. The target introduces a round-trip delay $\tau$. The derivation below is written for one heterodyne channel; the same treatment applies independently to the probe and reference channels. The returned probe field $E_p(t-\tau)$ and the local-oscillator (LO) field $E_{\mathrm{LO}}(t)$ are written as
\begin{equation}
E_p(t-\tau)=\sum_k P_k
e^{j\left[(\omega_0+k\Omega)(t-\tau)+\phi_p(t-\tau)\right]},
\end{equation}
\begin{equation}
E_{\mathrm{LO}}(t)=\sum_n L_n
e^{j\left[(\omega_0+\Delta\omega+n\Omega+n\Delta\Omega)t+\phi_{\mathrm{LO}}(t)\right]},
\end{equation}
where $\omega_0$ is the signal-comb optical carrier angular frequency, $\Omega$ is the signal-comb repetition angular frequency, $P_k$ and $L_n$ are the complex comb-tooth amplitudes, and $\phi_p(t)$ and $\phi_{\mathrm{LO}}(t)$ are the phase fluctuations of the two optical fields. After balanced photodetection, the detected radio-frequency (RF) signal is represented by the complex analytic envelope
\begin{equation}
\widetilde{i}(t)=\mathcal{R}_{\mathrm{PD}}E_{\mathrm{LO}}(t)E_p^*(t-\tau),
\end{equation}
where $\mathcal{R}_{\mathrm{PD}}$ is the effective responsivity of the balanced photodetector. Substitution of the two optical fields gives
\begin{align}
\widetilde{i}(t)=\mathcal{R}_{\mathrm{PD}}\sum_n\sum_k L_nP_k^*
&e^{j[\Delta\omega+(n-k)\Omega+n\Delta\Omega]t}\nonumber\\
&\times e^{j(\omega_0+k\Omega)\tau}
e^{j[\phi_{\mathrm{LO}}(t)-\phi_p(t-\tau)]}.
\end{align}
Only the low-frequency beat terms satisfying $n=k$ are retained in dual-comb detection. The complex RF-comb signal then becomes
\begin{align}
\widetilde{i}(t)
&=\mathcal{R}_{\mathrm{PD}}\sum_n L_nP_n^*
e^{j(\omega_0+n\Omega)\tau}
e^{j(\Delta\omega+n\Delta\Omega)t}
e^{j\eta(t)}\nonumber\\
&=e^{j[\Delta\omega t+\eta(t)]}
\sum_n B_n e^{jn\Delta\Omega t},
\end{align}
where
\begin{equation}
\eta(t)=\phi_{\mathrm{LO}}(t)-\phi_p(t-\tau)
\end{equation}
is the effective relative phase noise. It includes the accumulated phase arising from time-dependent variations of the optical frequency offset. The instantaneous angular centre frequency of the original RF comb is therefore
\begin{equation}
\omega_c(t)
=
\frac{\mathrm{d}}{\mathrm{d}t}
\left[\Delta\omega t+\eta(t)\right]
=
\Delta\omega+\frac{\mathrm{d}\eta(t)}{\mathrm{d}t},
\end{equation}
where $\omega_c(t)=2\pi f_c(t)$. The complex coefficient of the $n$th RF-comb tooth is
\begin{equation}
B_n=\mathcal{R}_{\mathrm{PD}}L_nP_n^* e^{j(\omega_0+n\Omega)\tau}.
\end{equation}

If each RF-comb tooth could be separated without crosstalk, its demodulated phase would be
\begin{equation}
\Theta_n(t)=\arg(B_n)+\eta(t)
=\arg(A_n)+(\omega_0+n\Omega)\tau+\eta(t),
\end{equation}
where $B_n=A_n e^{j(\omega_0+n\Omega)\tau}$ and $\arg(A_n)$ is a time-invariant system phase. A separate reference calibration could remove this term for absolute ranging. Without such calibration, the component of the static phase that varies linearly with comb-tooth index contributes a fixed distance offset. Because $\eta(t)$ is common to all RF-comb teeth, the phase slope is
\begin{equation}
\frac{\partial\Theta_n}{\partial n}
=
\frac{\partial\arg(A_n)}{\partial n}+\Omega\tau.
\end{equation}
Thus, in the ideal tooth-resolved limit, the common phase noise changes only the phase intercept and does not affect either the static system contribution or the distance-dependent slope. Changes in the fitted slope therefore track changes in delay because the system contribution remains fixed.

\subsection*{1.1 Finite-time extraction of the RF-comb coefficients}

In practice, the RF-comb is measured over a finite duration $T$, and the measured analytic signal also contains additive detection, electronic and quantization noise $v(t)$. The angular frequency used to extract the $n$th RF-comb tooth is
\begin{equation}
\omega_{\mathrm{RF},n}=\Delta\omega+n\Delta\Omega.
\end{equation}
Time-dependent deviations of the instantaneous RF-comb centre frequency from this extraction frequency are represented by the accumulated phase $\eta(t)$.
The complex coefficient extracted at this frequency is
\begin{align}
Z_n
&=\frac{1}{T}\int_0^T[\widetilde{i}(t)+v(t)]
e^{-j\omega_{\mathrm{RF},n}t}\,\mathrm{d}t\nonumber\\
&=\sum_k B_k\frac{1}{T}\int_0^T
e^{j\eta(t)}e^{j(k-n)\Delta\Omega t}\,\mathrm{d}t+W_n\nonumber\\
&=B_nQ_0+\sum_{k\ne n}B_kQ_{k-n}+W_n,
\end{align}
where
\begin{equation}
Q_\ell=\frac{1}{T}\int_0^T e^{j\eta(t)}
e^{j\ell\Delta\Omega t}\,\mathrm{d}t.
\end{equation}
Here,
\begin{equation}
W_n=\frac{1}{T}\int_0^T v(t)
e^{-j\omega_{\mathrm{RF},n}t}\,\mathrm{d}t
\end{equation}
is the additive-noise coefficient in the same extraction window. The first term is the desired RF-comb tooth, the summation describes phase-noise-induced mixing from neighbouring RF-comb teeth, and $W_n$ is the additive receiver-noise contribution.

In the absence of phase and additive noise, $\eta(t)=0$, $W_n=0$, and
\begin{align}
Z_n
&=B_n+\sum_{k\ne n}B_k
e^{j(k-n)\Delta\Omega T/2}
\frac{\sin[(k-n)\Delta\Omega T/2]}
{(k-n)\Delta\Omega T/2}.
\end{align}
If the measurement duration satisfies
\begin{equation}
T=N_{\mathrm{per}}\frac{2\pi}{\Delta\Omega}=\frac{N_{\mathrm{per}}}{\Delta f_r},
\qquad N_{\mathrm{per}}\in\mathbb{Z},
\end{equation}
where $\Delta f_r=\Delta\Omega/(2\pi)$, all cross terms vanish and
\begin{equation}
Z_n=B_n.
\end{equation}
Therefore, without phase noise, an acquisition window containing an integer number of RF-comb periods preserves the orthogonality of the RF-comb teeth and prevents finite-window spectral leakage.

\subsection*{1.2 Independent free-running dual-comb sources}

For two independent free-running comb sources, the phase fluctuations are uncorrelated and
\begin{equation}
\eta(t)=\phi_{\mathrm{LO}}(t)-\phi_p(t-\tau)
\end{equation}
can vary substantially during one acquisition window. The desired coherent term is
\begin{equation}
Q_0=\frac{1}{T}\int_0^T e^{j\eta(t)}\,\mathrm{d}t.
\end{equation}
When $T$ is shorter than the mutual coherence time, $\eta(t)$ changes slowly and $Q_0\approx e^{j\eta_0}$. When $T$ exceeds the mutual coherence time, coherent averaging reduces $|Q_0|$, which approaches zero in the strongly decorrelated limit.

For Lorentzian lasers with linewidths $\Delta\nu_1$ and $\Delta\nu_2$, the approximate relative linewidth is
\begin{equation}
\Delta\nu_{\mathrm{rel}}=\Delta\nu_1+\Delta\nu_2.
\end{equation}
The RF-comb power spectral density can then be approximated as
\begin{equation}
S_i(f)=\sum_n |B_n|^2L(f-f_n),
\end{equation}
where
\begin{equation}
f_n=\frac{\Delta\omega+n\Delta\Omega}{2\pi}
\end{equation}
gives the frequency position of the $n$th RF-comb tooth, and a normalized Lorentzian line shape is
\begin{equation}
L(f)=\frac{1}{\pi}\frac{\Delta\nu_{\mathrm{rel}}/2}
{f^2+(\Delta\nu_{\mathrm{rel}}/2)^2}.
\end{equation}
At the centre frequency $f_n$, the crosstalk ratio from the $(n+\ell)$th tooth can be estimated as
\begin{equation}
\rho_{n+\ell\rightarrow n}\approx
\left|\frac{B_{n+\ell}}{B_n}\right|^2
\frac{L(\ell\Delta f_r)}{L(0)}.
\end{equation}
For the Lorentzian line shape,
\begin{equation}
\frac{L(\ell\Delta f_r)}{L(0)}=
\frac{1}{1+\left(2\ell\Delta f_r/
\Delta\nu_{\mathrm{rel}}\right)^2}.
\end{equation}
Thus, a larger RF-comb spacing or a smaller relative laser linewidth reduces spectral overlap and the associated corruption of the extracted comb-tooth phases.

RF-tooth resolvability and probe--reference phase correlation represent two distinct consequences of laser phase noise. RF-tooth broadening is governed by the relative signal--LO linewidth compared with the RF-comb spacing. By contrast, probe--reference correlation depends on the phase increments sampled at the differential path delays. When the differential LO-path delay is negligible, the LO phase is largely common to the two channels, so this correlation is determined mainly by the signal-laser linewidth and the probe--reference signal-path imbalance. The combined linewidth used above to describe RF spectral broadening and the signal-laser coherence timescale used in the experiment therefore represent different physical constraints.

\subsection*{1.3 Propagation from extracted component-phase errors to distance error}

For either the probe or reference channel, $q\in\{p,r\}$, the finite-window coefficient can be factored as
\begin{equation}
Z_{q,n}
=B_{q,n}Q_{q,0}\left(1+\xi_{q,n}\right),
\end{equation}
where
\begin{align}
\xi_{q,n}
&=\xi_{q,n}^{\mathrm{PN}}+\xi_{q,n}^{\mathrm{add}},\nonumber\\
\xi_{q,n}^{\mathrm{PN}}
&=\frac{\displaystyle\sum_{k\ne n}B_{q,k}Q_{q,k-n}}
{B_{q,n}Q_{q,0}},\nonumber\\
\xi_{q,n}^{\mathrm{add}}
&=\frac{W_{q,n}}{B_{q,n}Q_{q,0}}.
\end{align}
Here, $\xi_{q,n}^{\mathrm{PN}}$ is the phase-noise-induced mixing contribution and $\xi_{q,n}^{\mathrm{add}}$ is the additive receiver-noise contribution. Provided that the desired coefficient $B_{q,n}Q_{q,0}$ is non-zero, consistent $2\pi$ phase unwrapping across the comb index gives the extracted component phase as
\begin{equation}
\arg Z_{q,n}
=\arg B_{q,n}+\arg Q_{q,0}+\epsilon_{q,n},
\end{equation}
with the exact residual phase error
\begin{equation}
\epsilon_{q,n}=\arg\left(1+\xi_{q,n}\right).
\end{equation}
When $|\xi_{q,n}|\ll1$,
\begin{equation}
\epsilon_{q,n}
\simeq
\operatorname{Im}\left(
\xi_{q,n}^{\mathrm{PN}}+\xi_{q,n}^{\mathrm{add}}
\right).
\end{equation}
Although $\arg Q_{q,0}$ is common to the extracted teeth within channel $q$ and therefore changes only the phase intercept, $\epsilon_{q,n}$ is generally comb-index dependent because the complex ratios $B_{q,k}/B_{q,n}$ differ with $n$.

After probe--reference phase subtraction, the extracted phase at index $n$ can be written as
\begin{equation}
\Delta\Theta_n
=b+\Delta\theta_{\mathrm{inst},n}
+n\Omega\Delta\tau
+\Delta\epsilon_n,
\qquad
\Delta\epsilon_n=\epsilon_{p,n}-\epsilon_{r,n},
\end{equation}
where $b$ contains terms independent of $n$, $\Delta\theta_{\mathrm{inst},n}$ is the static differential instrumental phase, and $\Delta\tau=\tau_p-\tau_r$. For a selected set of indices $\{n_i\}_{i=1}^{K}$, define the unweighted linear-fit weights
\begin{equation}
w_i=
\frac{n_i-\overline n}
{\displaystyle\sum_{j=1}^{K}(n_j-\overline n)^2},
\qquad
\overline n=\frac{1}{K}\sum_{i=1}^{K}n_i.
\end{equation}
These weights satisfy $\sum_iw_i=0$ and $\sum_iw_in_i=1$. The fitted phase slope is therefore
\begin{equation}
\widehat s_{\mathrm{raw}}
=\Omega\Delta\tau+s_{\mathrm{inst}}
+\sum_{i=1}^{K}w_i\Delta\epsilon_{n_i},
\end{equation}
where $s_{\mathrm{inst}}=\sum_iw_i\Delta\theta_{\mathrm{inst},n_i}$ is time invariant and contributes a fixed distance offset. Using $\Omega=2\pi f_{r,\mathrm{sig}}$, the retrieved distance becomes
\begin{equation}
\widehat d_{\mathrm{raw}}
=\frac{c\Delta\tau}{2}+d_0+\delta d_{\mathrm{raw}},
\end{equation}
with
\begin{equation}
\delta d_{\mathrm{raw}}
=\frac{c}{4\pi f_{r,\mathrm{sig}}}
\sum_{i=1}^{K}w_i
\left(\epsilon_{p,n_i}-\epsilon_{r,n_i}\right)
.
\end{equation}
Thus, common phase terms affect only the fitted intercept, whereas the projection of comb-index-dependent phase errors onto the linear-slope direction produces distance error. When the desired term no longer dominates, the first-order approximation fails; the exact phase expression must then be used, and phase slips, incorrect unwrapping or large errors in the retrieved distance may occur. Reduced probe--reference phase correlation can increase the residual differential phase error and, through its projection onto the fitted phase slope, the retrieved-distance error.

\subsection*{1.4 Self-conjugate phase-noise cancellation}

Conjugate phase-noise cancellation (CPNC) starts from the complex RF-comb envelope
\begin{equation}
\widetilde{i}(t)
=
e^{j[\Delta\omega t+\eta(t)]}
\sum_n B_n e^{jn\Delta\Omega t},
\end{equation}
with
\begin{equation}
B_n=A_n e^{j(\omega_0+n\Omega)\tau}.
\end{equation}
The signal is multiplied by its complex conjugate:
\begin{equation}
y(t)=\widetilde{i}(t)\widetilde{i}^*(t).
\end{equation}
Substitution gives
\begin{align}
y(t)
&=\left\{e^{j[\Delta\omega t+\eta(t)]}
\sum_nB_n e^{jn\Delta\Omega t}\right\}\nonumber\\
&\quad\times
\left\{e^{-j[\Delta\omega t+\eta(t)]}
\sum_kB_k^* e^{-jk\Delta\Omega t}\right\}\nonumber\\
&=\sum_n\sum_kB_nB_k^*e^{j(n-k)\Delta\Omega t}.
\end{align}
The complete tooth-common phase factor $e^{j[\Delta\omega t+\eta(t)]}$ is therefore cancelled exactly by its conjugate. Because the cancellation holds independently of the temporal form of $\eta(t)$, it includes the common phase evolution associated with RF-comb centre-frequency drift. Furthermore,
\begin{equation}
B_nB_k^*=A_nA_k^* e^{j(n-k)\Omega\tau},
\end{equation}
and hence
\begin{equation}
y(t)=\sum_n\sum_kA_nA_k^*
e^{j(n-k)\Omega\tau}
e^{j(n-k)\Delta\Omega t}.
\end{equation}
Defining the difference order $m=n-k$ gives
\begin{equation}
y(t)=\sum_m\left(\sum_kA_{k+m}A_k^*\right)
e^{jm\Omega\tau}e^{jm\Delta\Omega t}.
\end{equation}
With
\begin{equation}
C_m=\sum_kA_{k+m}A_k^*,
\end{equation}
the CPNC output is a difference-frequency comb
\begin{equation}
y(t)=\sum_m C_m e^{jm\Omega\tau}e^{jm\Delta\Omega t}.
\end{equation}
The phase of its $m$th difference-frequency component is
\begin{equation}
\Theta_m=\arg(C_m)+m\Omega\tau.
\end{equation}
The same self-conjugate operation was applied independently to the probe and reference channels. Denoting the corresponding delays by $\tau_p$ and $\tau_r$, subtracting their difference-frequency-component phases gives
\begin{equation}
\Phi_m
=\Theta_m^{(p)}-\Theta_m^{(r)}
=\phi_m^{\mathrm{inst}}
+m\Omega\Delta\tau,
\qquad
\Delta\tau=\tau_p-\tau_r,
\end{equation}
where $\phi_m^{\mathrm{inst}}$ is the static differential instrumental phase between the CPNC-processed probe and reference channels. Its approximately linear component contributes a fixed distance offset, whereas changes in the fitted slope track changes in the probe--reference delay. For a round-trip ranging geometry, the corresponding distance change is $\Delta d=c\Delta\tau/2$. A one-time reference measurement could additionally determine the fixed instrumental offset for calibrated absolute ranging. In the stability and relative-displacement measurements reported here, no independent instrumental-phase calibration was applied; the static response was treated as time invariant and therefore did not affect the reported distance fluctuations or displacement increments.

Importantly, CPNC is applied independently within the probe and reference channels. Each channel's tooth-common phase factor is cancelled before their difference-frequency-component phases are compared, so the cancellation does not require the two channel phases to remain correlated. Within the ideal tooth-common-phase model, a probe--reference path imbalance that weakens conventional phase-noise subtraction does not reintroduce the cancelled phase factor into the CPNC observable. Physical variations of the probe--reference delay are nevertheless retained because they change the distance-dependent inter-tooth phase slope.

The complete tooth-common phase evolution arising from laser phase noise and optical-frequency-offset drift cancels exactly from the ideal signal--signal product, but additive noise is not removed by the self-conjugate operation. With $\widetilde i_{\mathrm{meas}}(t)=\widetilde i(t)+v(t)$,
\begin{align}
y_{\mathrm{meas}}(t)
&=\widetilde i_{\mathrm{meas}}(t)
\widetilde i_{\mathrm{meas}}^*(t)\nonumber\\
&=\widetilde i(t)\widetilde i^*(t)
+\widetilde i(t)v^*(t)
+v(t)\widetilde i^*(t)
+v(t)v^*(t).
\end{align}
The middle two terms are first-order signal--noise contributions, and the last term is a second-order noise--noise contribution. Together with non-common component fluctuations and non-ideal receiver filtering, they are represented by a residual phase error $\epsilon_{q,m}^{\mathrm{CPNC}}$ in channel $q$. The measured probe--reference phase is therefore written as
\begin{equation}
\widehat{\Phi}_m
=\phi_m^{\mathrm{inst}}
+m\Omega\Delta\tau
+\Delta\epsilon_m^{\mathrm{CPNC}},
\qquad
\Delta\epsilon_m^{\mathrm{CPNC}}
=\epsilon_{p,m}^{\mathrm{CPNC}}
-\epsilon_{r,m}^{\mathrm{CPNC}}.
\end{equation}
Using the same unweighted-fit construction as above, with $n_i$ replaced by the selected difference orders $m_i$, gives
\begin{equation}
\delta d_{\mathrm{CPNC}}
=\frac{c}{4\pi f_{r,\mathrm{sig}}}
\sum_{i=1}^{K}w_i^{(m)}
\Delta\epsilon_{m_i}^{\mathrm{CPNC}}
,
\end{equation}
where
\begin{equation}
w_i^{(m)}
=\frac{m_i-\overline m}
{\displaystyle\sum_{j=1}^{K}(m_j-\overline m)^2}.
\end{equation}
This expression shows that CPNC removes the direct common-phase mixing pathway but does not imply a noise-free distance estimate.

To distinguish distance error from phase-fit residual, any residual differential phase-error vector can be decomposed as
\begin{equation}
\Delta\epsilon_m
=a+b(m-\overline m)+r_m,
\end{equation}
with $\sum_mr_m=0$ and $\sum_m(m-\overline m)r_m=0$. The constant term $a$ changes only the fitted intercept. The linear term changes the fitted distance by
\begin{equation}
\delta d_{\mathrm{slope}}
=\frac{c\,b}{4\pi f_{r,\mathrm{sig}}},
\end{equation}
without necessarily increasing the phase-fit RMSE, whereas $r_m$ contributes to the non-linear fit residual and hence to the operational phase-fit success criterion. Temporal variations of the slope-like term $b$, together with physical probe--reference path fluctuations, determine the measured distance fluctuation and averaging behaviour. Variations of $r_m$ govern fit quality and, outside the linear-error regime, can contribute to phase-unwrapping failures and erroneous distance estimates. Quantitative prediction of these metrics requires noise amplitudes and temporal correlations beyond this first-order propagation model.

The derivation therefore shows that common laser phase noise and optical-frequency-offset drift do not intrinsically alter the distance-dependent phase slope, but finite-window mixing of broadened RF-comb teeth converts the corresponding tooth-common phase evolution into comb-index-dependent phase errors that bias the retrieved distance. By cancelling the complete tooth-common phase factor before the complex coefficients are extracted and their phases evaluated, CPNC suppresses this error pathway while preserving the distance-dependent phase slope. Residual distance errors can nevertheless arise from additive receiver noise and non-common component fluctuations through their projection onto the fitted phase slope.

\clearpage
\suppnote{2}{Complex spectral-component extraction, phase-slope retrieval and performance metrics}
\begingroup
\setlength{\abovedisplayskip}{6pt plus 2pt minus 2pt}
\setlength{\belowdisplayskip}{6pt plus 2pt minus 2pt}
\setlength{\abovedisplayshortskip}{3pt plus 2pt}
\setlength{\belowdisplayshortskip}{5pt plus 2pt minus 2pt}

\subsection*{2.1 Complex spectral-component extraction and distance retrieval}

The raw and CPNC branches were evaluated from the same recorded probe and reference traces. For the $s$th segment and channel $q\in\{p,r\}$, the complex coefficient at a target frequency $f_k^{(s)}$ was evaluated directly using a single-frequency discrete Fourier sum,
\begin{equation}
X_{q,k}^{(s)}
=
\Delta t\sum_{l=0}^{L-1}x_q^{(s)}(t_l)
e^{-j2\pi f_k^{(s)}t_l},
\end{equation}
where $L$ is the number of samples in one segment, $\Delta t$ is the sampling interval and $x_q^{(s)}$ denotes the branch-specific signal. Here, spectral-component extraction refers to the direct evaluation of the complex coefficient at each selected frequency, from which the corresponding phase is obtained. No rounding to the nearest frequency on a uniform Fourier grid was required.

For the raw-processing branch, the target frequency of RF-comb tooth $n$ in segment $s$ was
\begin{equation}
f_{n,\mathrm{raw}}^{(s)}
=
f_c^{(s)}+n\Delta f_r,
\end{equation}
where $f_c^{(s)}$ is the local RF-comb centre frequency obtained from the auxiliary laser-beat channel as described in Supplementary Note 3. For the CPNC branch, the complete probe and reference waveforms were first converted into analytic signals by a Hilbert transform and then multiplied pointwise by their complex conjugates. The resulting difference-frequency components occurred at the fixed frequencies
\begin{equation}
f_{m,\mathrm{CPNC}}=m\Delta f_r,
\end{equation}
and were extracted without using the auxiliary laser-beat channel or a monitor-derived RF-comb centre-frequency estimate. For the static-ranging and linewidth-comparison measurements, 15 RF-comb teeth were used in the raw-processing branch and 15 difference-frequency components in the CPNC branch. The corresponding numbers were 12 and 12 for the stepwise displacement measurements.

At each selected frequency, the probe--reference differential phase was calculated as
\begin{equation}
\Delta\phi_k^{(s)}
=
\operatorname{unwrap}_k
\left\{
\arg\left[X_{p,k}^{(s)}X_{r,k}^{(s)*}\right]
\right\},
\end{equation}
where phase unwrapping was performed along the ordered frequency components. An unweighted linear least-squares fit,
\begin{equation}
\Delta\phi_k^{(s)}=\beta_s f_k^{(s)}+b_s,
\end{equation}
then provided the phase slope $\beta_s$. If the preliminary fit fell on the adjacent negative-distance branch, a linear phase ramp of
\begin{equation}
2\pi\frac{f_k^{(s)}-\overline{f}^{(s)}}{\Delta f_r}
\end{equation}
was added before refitting. This operation shifts the retrieved distance by one non-ambiguity range without changing the phase residuals. The distance for segment $s$ was obtained from
\begin{equation}
d_s
=
\frac{c\,\beta_s\Delta f_r}{4\pi f_{r,\mathrm{sig}}},
\end{equation}
where $f_{r,\mathrm{sig}}=25.000~\mathrm{GHz}$ is the signal-comb repetition rate. When non-ambiguity-range wrapping was required, the distance was mapped to $[0,c/(2f_{r,\mathrm{sig}}))$.

\subsection*{2.2 Numerical evaluation of performance metrics}

The phase-fit quality of each segment was quantified from the residuals
\begin{equation}
r_k^{(s)}
=
\Delta\phi_k^{(s)}-
\left(\beta_s f_k^{(s)}+b_s\right)
\end{equation}
using the root-mean-square error (RMSE),
\begin{equation}
\mathrm{RMSE}_{\phi}^{(s)}
=
\sqrt{\frac{1}{K}\sum_{k=1}^{K}\left[r_k^{(s)}\right]^2},
\end{equation}
where $K$ is the number of fitted components. A segment was counted as successful when $\mathrm{RMSE}_{\phi}^{(s)}<1~\mathrm{rad}$. The phase-fit success rate was calculated as $N_{\mathrm{success}}/N_{\mathrm{total}}$. The $1~\mathrm{rad}$ threshold was used as a fixed operational criterion because it is substantially smaller than $\pi$, thereby identifying frames whose phase samples remained concentrated around a single linear phase-slope model. It is not intended as a universal physical boundary for successful ranging. The same criterion was applied to all linewidth configurations and both processing branches. The coefficient of determination was also recorded as a diagnostic quantity but was not included in the success criterion.

The single-frame distance fluctuation was reported as the sample standard deviation of the complete distance sequence,
\begin{equation}
s_d
=
\sqrt{\frac{1}{N_{\mathrm{seg}}-1}
\sum_{s=1}^{N_{\mathrm{seg}}}
\left(d_s-\overline{d}\right)^2}.
\end{equation}
The distance standard deviations and Allan deviations reported in the main text were calculated from all segmented frames without RMSE-based rejection. The $1~\mathrm{rad}$ threshold was used only to calculate the phase-fit success rate and did not filter the reported distance sequences.

The overlapping Allan deviation was calculated directly from the distance sequence. For an averaging factor $L_{\mathrm{A}}$, the local average was
\begin{equation}
\overline{d}_i^{(L_{\mathrm{A}})}
=
\frac{1}{L_{\mathrm{A}}}
\sum_{s=i}^{i+L_{\mathrm{A}}-1}d_s,
\end{equation}
and the corresponding Allan deviation was
\begin{equation}
\sigma_{\mathrm{A}}(\tau)
=
\sqrt{
\frac{1}{2(N_{\mathrm{seg}}-2L_{\mathrm{A}}+1)}
\sum_{i=1}^{N_{\mathrm{seg}}-2L_{\mathrm{A}}+1}
\left[
\overline{d}_{i+L_{\mathrm{A}}}^{(L_{\mathrm{A}})}-
\overline{d}_i^{(L_{\mathrm{A}})}
\right]^2
},
\qquad
\tau=L_{\mathrm{A}}T_{\mathrm{seg}}.
\end{equation}
Here, $N_{\mathrm{seg}}$ is the total number of segmented frames. Only averaging factors satisfying $L_{\mathrm{A}}\leq\lfloor N_{\mathrm{seg}}/2\rfloor$ were evaluated.

\endgroup

\clearpage
\suppnote{3}{Auxiliary laser-beat channel}

Because the two seed lasers are free-running, their optical frequency offset is not actively locked and drifts during data acquisition. The self-conjugate CPNC operation removes the tooth-common phase evolution associated with this drift from the resulting difference-frequency comb. Although CPNC does not require the instantaneous optical frequency offset, the raw-processing branch uses the corresponding RF-comb centre frequency to locate the RF-comb teeth and maintain their correct assignment during segment-wise processing. An auxiliary laser-beat channel was therefore implemented to track the optical frequency offset between the two seed lasers.

\begin{figure}[htbp]
  \centering
  \includegraphics[width=0.68\textwidth]{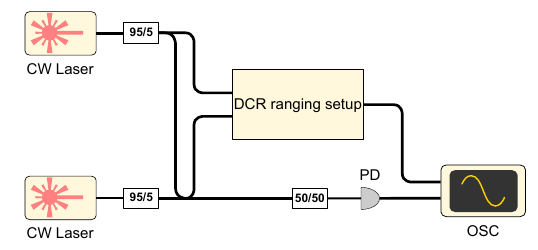}
  \caption{Auxiliary laser-beat channel for optical-frequency-offset tracking. Five percent of the optical power from each free-running CW seed laser is tapped by a 95/5 fiber coupler, while the remaining power is directed to the dual-comb ranging setup. The tapped fields are combined by a 50/50 fiber coupler and detected by a single photodetector (XPDV3120, Finisar). The resulting electrical beat note is recorded by the oscilloscope and used to determine the local RF-comb centre frequency during offline processing. The auxiliary laser-beat signal is used only for RF-comb centre-frequency tracking and RF-comb tooth assignment in the raw-processing branch; it is not used for CPNC processing or CPNC-based distance retrieval. CPNC, conjugate phase-noise cancellation; CW, continuous-wave; RF, radio-frequency.}
  \label{fig:supp-monitor}
\end{figure}

The auxiliary laser-beat channel is shown in Fig.~\ref{fig:supp-monitor}. Five percent of the optical power from each continuous-wave (CW) seed laser is tapped using a 95/5 fiber coupler, while the remaining 95\% is directed to the dual-comb ranging setup. The two tapped optical fields are combined by a 50/50 fiber coupler and detected by a single photodetector (XPDV3120, Finisar). The resulting electrical beat frequency,
\begin{equation}
f_{\mathrm{beat}}(t)
=
\left|f_{\mathrm{sig}}(t)-f_{\mathrm{LO}}(t)\right|,
\end{equation}
represents the instantaneous optical frequency difference between the signal-comb and LO-comb seed lasers. The auxiliary laser-beat signal is recorded by the same oscilloscope together with the probe and reference RF-comb signals.

During offline processing, the recorded traces are divided into segments of duration $T_{\mathrm{seg}}=1/\Delta f_r$, where $\Delta f_r=30~\mathrm{MHz}$. Ten consecutive segments are grouped into one optical-frequency-offset tracking window. Within each window, the dominant peak in the Fourier spectrum of the auxiliary laser-beat channel is used to estimate the local optical frequency offset and the corresponding RF-comb centre frequency, which is then assigned to the corresponding probe and reference segments for RF-comb tooth localization and extraction.

The auxiliary laser-beat channel is used exclusively for offset tracking and RF-comb tooth assignment in the raw-processing branch. It is not used in the CPNC branch or in CPNC-based distance retrieval.

\clearpage
\suppnote{4}{Seed-laser linewidth characterization}

\begin{figure}[htbp]
  \centering
  \includegraphics[width=0.90\textwidth]{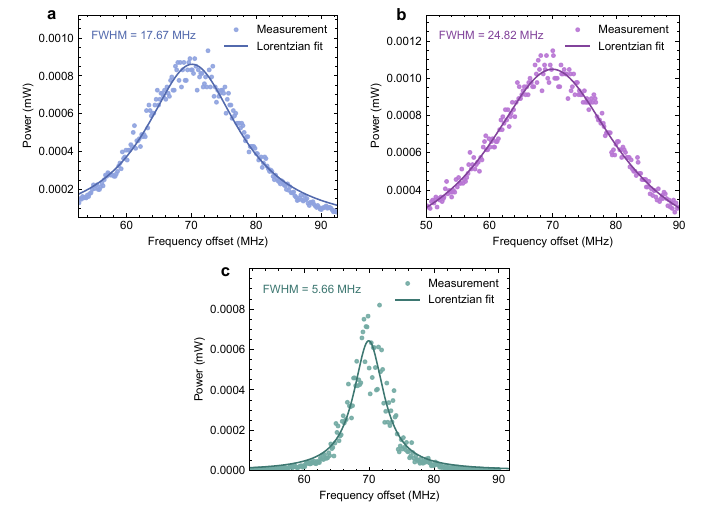}
  \caption{Delayed self-heterodyne linewidth measurements of the three MHz-linewidth DFB seed lasers. \textbf{a}, Measured electrical beat spectrum and Lorentzian fit for the $9~\mathrm{MHz}$ laser, giving a beat-note FWHM of $17.67~\mathrm{MHz}$ and an inferred optical linewidth of $8.83~\mathrm{MHz}$. \textbf{b}, Corresponding measurement for the $12~\mathrm{MHz}$ laser, giving a beat-note FWHM of $24.82~\mathrm{MHz}$ and an inferred optical linewidth of $12.41~\mathrm{MHz}$. \textbf{c}, Corresponding measurement for the $3~\mathrm{MHz}$ laser, giving a beat-note FWHM of $5.66~\mathrm{MHz}$ and an inferred optical linewidth of $2.83~\mathrm{MHz}$. Markers show the measured spectra, and solid lines show the Lorentzian fits. DFB, distributed-feedback; FWHM, full width at half maximum.}
  \label{fig:supp-linewidth}
\end{figure}

The linewidths of the three free-running distributed-feedback (DFB) seed lasers used in the MHz-linewidth configurations were characterized by delayed self-heterodyne interferometry (DSHI) [1]. The measured electrical beat spectra were fitted with Lorentzian profiles. Under the Lorentzian fitting convention used here, the optical linewidth was taken as half the fitted beat-note full width at half maximum (FWHM). The fitted beat-note FWHM values were $17.67~\mathrm{MHz}$, $24.82~\mathrm{MHz}$ and $5.66~\mathrm{MHz}$ for the lasers referred to as the $9~\mathrm{MHz}$, $12~\mathrm{MHz}$ and $3~\mathrm{MHz}$ sources, respectively. These values correspond to inferred optical linewidths of $8.83~\mathrm{MHz}$, $12.41~\mathrm{MHz}$ and $2.83~\mathrm{MHz}$ (Fig.~\ref{fig:supp-linewidth}a--c). The rounded linewidths are used throughout the main text and Supplementary Information.

\clearpage
\suppnote{5}{Influence of fitted component number and spectral span}

\subsection*{5.1 Reference precision scaling with fitted component number}

The phase-slope precision in frequency-domain dual-comb ranging depends on the number, span and noise statistics of the fitted phase samples. For conventional comb-line fitting, Wang et al. derived the ideal distance precision [2]
\begin{equation}
\sigma_{d,\mathrm{Wang}}
=\frac{
\sqrt{3(1+\alpha)}\,c
}{
2\pi B\sqrt{N_{\mathrm{line}}}
\sqrt{\mathrm{rSNR}_{\mathrm{eff}}}
},
\end{equation}
where $N_{\mathrm{line}}$ is the number of fitted comb lines, $B\simeq N_{\mathrm{line}}f_r$ is the used optical comb bandwidth in the large-$N_{\mathrm{line}}$ approximation, $\mathrm{rSNR}_{\mathrm{eff}}$ is the effective RF spectral signal-to-noise ratio and $\alpha$ accounts for the relative noise contributions of the signal and reference arms. This expression gives an $N_{\mathrm{line}}^{-1/2}$ improvement when $B$ is fixed. When the line spacing is fixed so that $B$ increases with $N_{\mathrm{line}}$, the ideal trend becomes $N_{\mathrm{line}}^{-3/2}$, provided that the effective phase noise does not increase.

Here, $N_{\mathrm{line}}$ counts the original optical comb lines used in conventional phase-slope fitting and should not be identified directly with $K$, which below denotes the number of signed CPNC difference-frequency components.

Because the CPNC output is real, its positive and negative difference orders form conjugate pairs. Let $M$ denote the number of non-redundant non-zero positive orders, so that the complete symmetric set contains $K=2M+1$ signed components, including the zero-order component. To obtain a reference scaling rather than a physical noise model for CPNC, we consider the idealized case of independent, equal-variance positive-order residual phase errors,
\begin{equation}
\boldsymbol{\Sigma}_{+}
=\sigma_{\Phi,+}^{2}\boldsymbol{I},
\end{equation}
where $\boldsymbol{\Sigma}_{+}$ is the covariance of the positive-order differential phases and $\sigma_{\Phi,+}$ is their common standard deviation. Applying the phase-error-to-distance relation derived in Supplementary Note 1 to the symmetric CPNC difference orders gives
\begin{equation}
\delta d
=\frac{c}{4\pi f_{r,\mathrm{sig}}}
\frac{\displaystyle\sum_{m=1}^{M}m\epsilon_m}
{\displaystyle\sum_{m=1}^{M}m^2},
\end{equation}
where $\epsilon_m\equiv\Delta\epsilon_m^{\mathrm{CPNC}}$ is the residual probe--reference phase error of positive difference order $m$. Using
\begin{equation}
\sum_{m=1}^{M}m^2
=\frac{M(M+1)(2M+1)}{6},
\end{equation}
the corresponding distance standard deviation is
\begin{equation}
\sigma_d
=\frac{c\,\sigma_{\Phi,+}}
{4\pi f_{r,\mathrm{sig}}}
\sqrt{
\frac{6}{M(M+1)(2M+1)}
}.
\end{equation}
Equivalently, in terms of the number of signed fitted components,
\begin{equation}
\sigma_d
=\frac{
\sqrt{6}\,c\,\sigma_{\Phi,+}
}{
2\pi f_{r,\mathrm{sig}}
\sqrt{K(K^2-1)}
},
\qquad K=2M+1.
\end{equation}
The large-$M$ limit of this statistical benchmark is $M^{-3/2}$. This expression is included only as a reference under the assumptions of independent, equal-variance positive-order phase errors. It is not a general physical scaling law for the present CPNC measurement because the component number, fitted frequency span, component amplitudes and phase-noise statistics change together as the fitted subset is expanded. In particular, the phase-noise-induced mixing and additive-noise contributions identified in Supplementary Note 1 can produce order-dependent and correlated residual phase errors that are not represented by the equal-variance benchmark above. The experimental dependence is therefore evaluated directly below rather than assigned a universal power law.

\subsection*{5.2 Experimental component-number and span analysis}

To examine the combined influence of fitted component number and spectral span on the present CPNC-processed measurement, we reprocessed the same $12~\mathrm{MHz}+9~\mathrm{MHz}$ data using nested, contiguous subsets of the 15 signed difference-frequency components. After ordering the components by frequency, each subset was centred on the zero-order component and expanded symmetrically by adding one conjugate pair at a time. The tested component numbers were $K=3,5,7,9,11,13$ and $15$, corresponding to $M=1,2,3,4,5,6$ and $7$ non-redundant non-zero positive orders and full signed RF fitting spans of $60$, $120$, $180$, $240$, $300$, $360$ and $420~\mathrm{MHz}$, respectively. Because the component number and fitting span were varied together, this analysis evaluates their combined influence rather than isolating the effect of component number alone. All subsets used the same segmented frames, CPNC processing, unweighted phase-slope fitting and distance-conversion procedure as in the main analysis. The phase-fit RMSE threshold of $1~\mathrm{rad}$ was used only to calculate the phase-fit success rate.

\begin{figure}[htbp]
  \centering
  \includegraphics[width=0.90\textwidth]{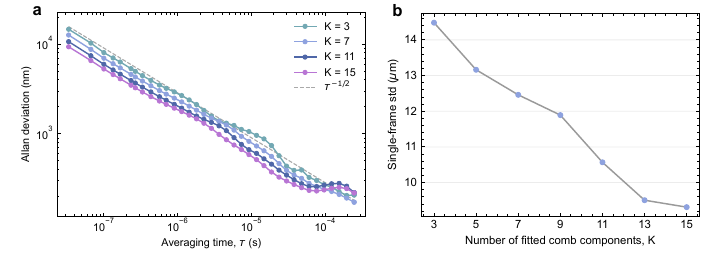}
  \caption{Influence of fitted component number and spectral span after CPNC. \textbf{a}, Allan deviation of the retrieved distance for nested component subsets selected from the 15 signed CPNC difference-frequency components in the $12~\mathrm{MHz}+9~\mathrm{MHz}$ measurement. Here, $K=2M+1$ counts the signed fitted components, corresponding to $M$ non-redundant non-zero positive orders, their conjugate negative orders and the zero-order component. Representative curves for $K=3,7,11$ and $15$ are shown for clarity. The dashed line indicates a $\tau^{-1/2}$ reference slope. Larger component numbers reduce the Allan deviation at short averaging times, where the curves approximately follow the $\tau^{-1/2}$ trend expected for white-noise-dominated averaging, whereas the curves converge and cross at longer averaging times. \textbf{b}, Single-frame distance standard deviation as a function of $K$. The full signed RF fitting span increases from $60~\mathrm{MHz}$ at $K=3$ to $420~\mathrm{MHz}$ at $K=15$. The standard deviation decreases monotonically from $14.48~\mu\mathrm{m}$ to $9.31~\mu\mathrm{m}$. All distance statistics were calculated using the same segmented frames and processing procedure. The phase-fit success rate was $100\%$ for all tested $K$ values; this operational rate was not used to infer component-number scaling. CPNC, conjugate phase-noise cancellation; RF, radio-frequency.}
  \label{fig:supp-diffK}
\end{figure}

Increasing $K$ reduced the single-frame distance standard deviation monotonically from $14.48~\mu\mathrm{m}$ at $K=3$ to $9.31~\mu\mathrm{m}$ at $K=15$ (Fig.~\ref{fig:supp-diffK}). The $K=15$ result reproduces the single-frame standard deviation reported in the main text for the full set of valid difference-frequency components.

All frames satisfied the $1~\mathrm{rad}$ phase-fit RMSE criterion for every tested $K$. This is reported as an operational outcome rather than as evidence for component-number scaling. In particular, for $K=3$ ($M=1$), conjugate symmetry constrains the three phase points at $m=-1,0$ and $1$ to be exactly collinear, so their phase-fit RMSE is zero by construction. The success-rate result was not used to determine the precision trend.

The Allan deviation curves show a complementary trend. At short averaging times, where the curves approximately follow the $\tau^{-1/2}$ behaviour expected for white-noise-dominated averaging, larger $K$ generally gives lower Allan deviation, consistent with improved phase-slope estimation over a broader fitted frequency span. At longer averaging times, the curves gradually converge and cross because the measurement is increasingly influenced by slower path and system fluctuations. The observed dependence therefore reflects the combined effects of fitted component number, frequency span and component signal-to-noise ratio, rather than a universal component-number scaling law. Accordingly, the ordering of the Allan deviations at $246~\mu\mathrm{s}$ should not be used to identify an optimum $K$. For the full $K=15$ set used for the main static-ranging result, the Allan deviation is $219~\mathrm{nm}$ at $246~\mu\mathrm{s}$.

\clearpage
\suppnote{6}{Detailed ranging results for mixed-linewidth laser configurations}

In the main text, we compare three CPNC-processed megahertz-linewidth configurations with the $100~\mathrm{kHz}+100~\mathrm{kHz}$ narrow-linewidth reference without CPNC. In each configuration, the first and second values denote the linewidths of the signal-comb and LO-comb seed lasers, respectively. Here, we present the detailed Allan deviation curves and single-frame distance sequences for the two mixed-linewidth configurations, $3~\mathrm{MHz}+100~\mathrm{kHz}$ and $9~\mathrm{MHz}+100~\mathrm{kHz}$.

For the $3~\mathrm{MHz}+100~\mathrm{kHz}$ configuration, the raw distance retrieval without CPNC exhibits strong instability. Its minimum Allan deviation is $2162~\mathrm{nm}$ at an averaging time of $195~\mu\mathrm{s}$, whereas the CPNC-processed measurement reaches $167~\mathrm{nm}$ at $47.7~\mu\mathrm{s}$ (Fig.~\ref{fig:supp-allan-3m}). The corresponding single-frame distance sequence also shows a substantial improvement (Fig.~\ref{fig:supp-distance-3m}). Without CPNC, the retrieved distances exhibit pronounced frame-to-frame fluctuations and intermittent outliers, resulting in a standard deviation of $168~\mu\mathrm{m}$. After CPNC, the distances become tightly concentrated around a single value, with a standard deviation of $4.89~\mu\mathrm{m}$. This represents an approximately 34-fold reduction in single-frame distance fluctuation.

For the $9~\mathrm{MHz}+100~\mathrm{kHz}$ configuration, the raw measurement is similarly degraded by laser-phase-noise-induced instability. Its minimum Allan deviation is $1603~\mathrm{nm}$ at $195~\mu\mathrm{s}$, whereas the CPNC-processed measurement reaches $195~\mathrm{nm}$ at $37.7~\mu\mathrm{s}$ (Fig.~\ref{fig:supp-allan-9m}). The raw distance sequence exhibits large fluctuations and occasional distance jumps, with a standard deviation of $103~\mu\mathrm{m}$ (Fig.~\ref{fig:supp-distance-9m}). After CPNC, the distance sequence remains stable over 1000 segmented frames, and its standard deviation is reduced to $4.52~\mu\mathrm{m}$, corresponding to an approximately 23-fold reduction.

Together, these results show that the CPNC improvement persists across both mixed-linewidth configurations, with minimum Allan deviations below $200~\mathrm{nm}$ and single-frame standard deviations below $5~\mu\mathrm{m}$.

\clearpage
\captionsetup{font=footnotesize,labelfont=bf,labelsep=period,hypcap=false}
\noindent
\begin{minipage}{\textwidth}
  \centering
  \includegraphics[width=0.62\linewidth]{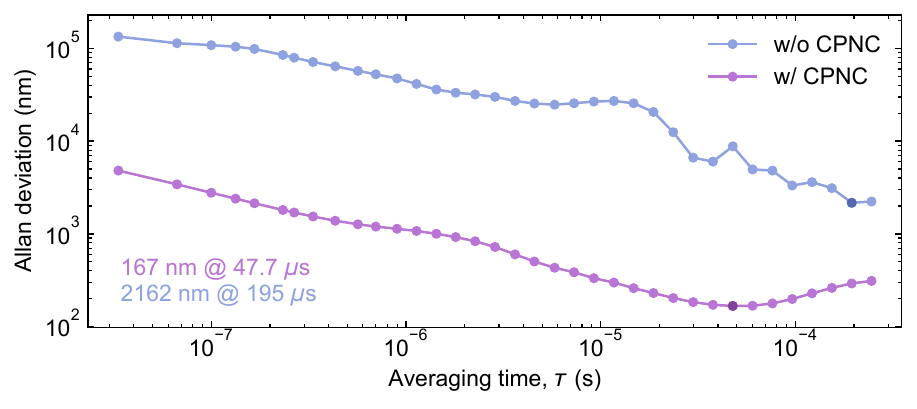}
  \captionof{figure}{Allan deviation for the $3~\mathrm{MHz}+100~\mathrm{kHz}$ configuration. The minimum is reduced from $2162~\mathrm{nm}$ to $167~\mathrm{nm}$ after CPNC. CPNC, conjugate phase-noise cancellation.}
  \label{fig:supp-allan-3m}
\end{minipage}

\vspace{0.15cm}

\noindent
\begin{minipage}{\textwidth}
  \centering
  \includegraphics[width=0.62\linewidth]{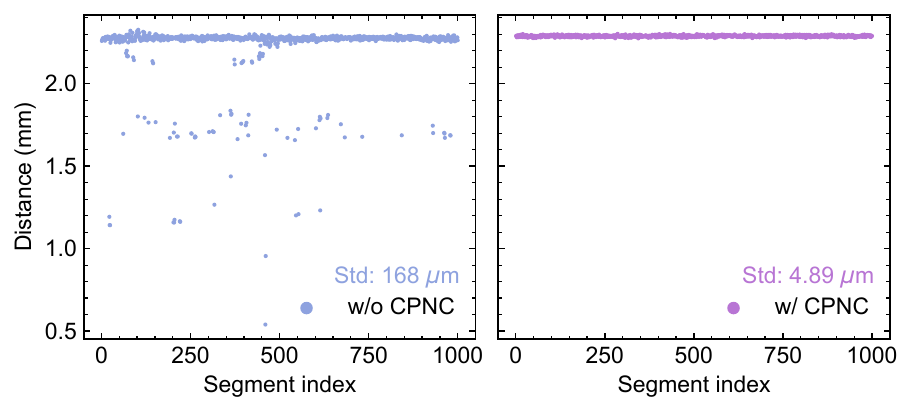}
  \captionof{figure}{Distance sequences over 1000 segmented frames for the $3~\mathrm{MHz}+100~\mathrm{kHz}$ configuration. The standard deviation is reduced from $168~\mu\mathrm{m}$ to $4.89~\mu\mathrm{m}$ after CPNC. CPNC, conjugate phase-noise cancellation.}
  \label{fig:supp-distance-3m}
\end{minipage}

\vspace{0.15cm}

\noindent
\begin{minipage}{\textwidth}
  \centering
  \includegraphics[width=0.62\linewidth]{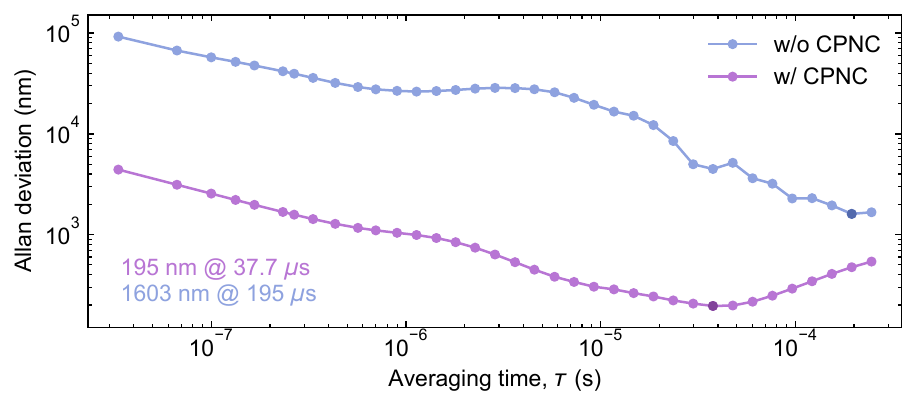}
  \captionof{figure}{Allan deviation for the $9~\mathrm{MHz}+100~\mathrm{kHz}$ configuration. The minimum is reduced from $1603~\mathrm{nm}$ to $195~\mathrm{nm}$ after CPNC. CPNC, conjugate phase-noise cancellation.}
  \label{fig:supp-allan-9m}
\end{minipage}

\vspace{0.15cm}

\noindent
\begin{minipage}{\textwidth}
  \centering
  \includegraphics[width=0.62\linewidth]{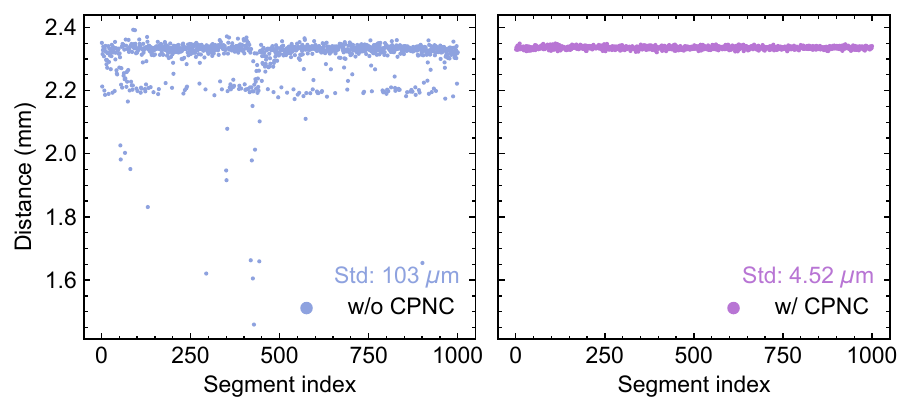}
  \captionof{figure}{Distance sequences over 1000 segmented frames for the $9~\mathrm{MHz}+100~\mathrm{kHz}$ configuration. The standard deviation is reduced from $103~\mu\mathrm{m}$ to $4.52~\mu\mathrm{m}$ after CPNC. CPNC, conjugate phase-noise cancellation.}
  \label{fig:supp-distance-9m}
\end{minipage}
\captionsetup{font=small,labelfont=bf,labelsep=period,hypcap=true}

\enlargethispage{3\baselineskip}
\renewcommand{\refname}{Supplementary References}
\phantomsection
\label{supprefs}

\end{document}